\documentstyle[12pt,amsfonts]{article}

\def\half{\textstyle{\frac{1}{2}}}
\def\quarter{\textstyle{\frac{1}{4}}}
\def\halfnu{\textstyle{\frac{1}{2\nu}}}

\def\H{{\cal H}}
\def\D{{\cal D}}
\def\De{\Delta}
\def\tr{{\rm tr}}
\def\E{{\rm I}\hskip-.2em{\rm E}}
\def\ra{\rightarrow}
\def\tint{{\textstyle\int}}
\def\hg{{\hat g}}
\def\hp{{\hat\pi}}
\def\hph{{\hat\phi}}
\def\s{\hskip.08em}
\def\d{\partial}
\def\o{\overline}
\def\a{\alpha}
\def\b{\begin{eqnarray*}}     
\def\e{\end{eqnarray*}}       
\def\bn{\begin{eqnarray}}     
\def\en{\end{eqnarray}}       
\def\<{\langle}
\def\>{\rangle}

\def\no{\nonumber}

\def\{{\lbrace}
\def\}{\rbrace}

\begin{document}
\title{Noncanonical quantization of gravity. II.\\
   Constraints and the physical Hilbert space}
\author{John R. Klauder\footnote{Electronic mail: klauder@phys.ufl.edu}\\
Departments of Physics and Mathematics\\
University of Florida\\
Gainesville, FL  32611}
\date{}
\maketitle
\begin{abstract}
The program of quantizing the gravitational field with the
help of affine field 
variables is continued. For completeness, a review of the selection 
criteria that singles out the affine fields, the alternative treatment 
of constraints, and the choice of the initial (before imposition of 
the constraints) ultralocal representation of the field operators 
is initially presented. As analogous examples demonstrate, the 
introduction and enforcement of the gravitational constraints will 
cause sufficient changes in the operator representations so that all 
vestiges of the initial ultralocal field operator representation 
disappear. To achieve this introduction and enforcement of the 
constraints, a well characterized phase space functional integral 
representation for the reproducing kernel of a suitably regularized 
physical Hilbert space is developed and extensively analyzed. 
\end{abstract}
\vfill\eject
\section{\sc INTRODUCTION AND THE \\MAIN POINTS OF PAPER I}
In a previous paper \cite{P-I} (hereafter referred to as P-I) an 
introduction and outline of a program of noncanonical quantization 
of the gravitational field was presented. The key concepts in the 
present approach are (i) a careful selection of the basic kinematical 
variables, (ii) the use of a quantization procedure that treats all 
constraints alike, (iii) the use of ultralocal field operator 
representations prior to introducing constraints, and (iv) the 
imposition of the gravitational constraints, a process in which all 
traces of the temporary ultralocal representation characteristics are 
replaced with the physically relevant ones. In this section, we briefly 
review topics (i), (ii), and (iii) which have largely been discussed 
already in P-I.   

\subsection{Nature of the basic gravitational variables}
One of the central requirements for the present program is the preservation, 
on quantization, of the positive-definite character of the spatial part 
of the classical metric $g_{ab}(x)$, $a,b\in\{1,2,3\}$ (or more 
generally $a,b\in\{1,\ldots,s\}$ in an $s$-dimensional space, $s\ge1$). 
Insisting on this requirement leads us to adopt {\it affine commutation 
relations} (in contrast, for example, to canonical commutation relations) 
which are expressed in terms of the local metric field operators 
$\hg_{ab}(x)$ $ [=\hg_{ba}(x)]$ (which were denoted by 
$\sigma_{ab}(x)$ in P-I) and suitable local ``scale'' field operators 
$\hp^c_d(x)$ (which were denoted by $\kappa^c_d(x)$ in P-I). The 
word ``local'' here is intended to mean that these expressions only 
become operators after smearing with suitable spatial test functions. 
In units where $\hbar=1$, which are commonly assumed throughout this 
paper, the basic set of affine commutation relations reads
\bn 
  &&\hskip.2cm[\hp^a_b(x),\,\hp^c_d(y)]=\half\s i\s[\s
\delta^c_b\hp^a_d(x)-\delta^a_d\hp^c_b(x)\s]\,\delta(x,y)\;,\no\\
  &&\hskip.1cm[\hg_{ab}(x),\,\hp^c_d(y)]=\half\s i\s[\s
\delta^c_a\hg_{bd}(x)+\delta^c_b\hg_{ad}(x)\s]\,\delta(x,y)\;,\label{afc}\\
&&[\hg_{ab}(x),\,\hg_{cd}(y)]=0\;.   \no \en
These commutation relations are translations of identical Poisson 
brackets (modulo $i\hbar$, of course) for corresponding classical 
fields, namely, the spatial metric $g_{ab}(x)$ and the mixed-valence 
(``scale'') field $\pi^c_d(x)\equiv g_{bd}(x)\pi^{bc}(x)$, along with 
the usual Poisson brackets between the metric field $g_{ab}(x)$ and the 
canonical momentum field $\pi^{cd}(x)$. While
{\it classically} there is essentially nothing to be gained by using the 
field
$\pi^c_d(x)$ rather than $\pi^{cd}(x)$, {\it quantum mechanically} the 
situation changes completely. This change arises because the affine 
commutation relations---and {\it only} the affine commutation 
relations---admit local self-adjoint operator solutions for both 
$\hg_{ab}(x)$ and $\hp^c_d(x)$ which in addition have the property 
that $\hg_{ab}(x)>0$ for all $x$. This latter property means that 
for any nonvanishing set $\{u^a\}$ of real numbers and any nonvanishing, 
nonnegative test function, $f(x)\ge0$, that
  \bn  \tint f(x)\s u^a\s\hg_{ab}(x)\s u^b\s d^3\!x>0\;.  \en 
Other choices of basic variables fail this test. For example, self-adjoint 
canonical variables lead to metrics that have spectra unbounded below as 
well as
above, while triad fields and their canonical partners lead to metrics 
that are nonnegative, but not necessarily positive definite.

\subsection{How quantum constraints are to be imposed}
Since gravity is a reparametrization invariant theory, it follows that 
the dynamics---indeed, {\it the entire physical content of the 
gravitational field}---enters through imposition of the relevant 
constraints, specifically, the diffeomorphism and Hamiltonian 
constraints \cite{ADM,DIRAC}. These constraints lead to an open set of 
{\it first-class} constraints in the classical theory, but on quantization 
and in virtue of an anomaly (or, alternatively, a factor ordering problem), 
they give rise to a set of operator constraints that, partially at least, 
are {\it second class} in nature. There exist several methods to deal with 
the quantum theory of second-class constraints in the literature, but 
generally such methods treat the first- and second-class constraints, 
and thereby the variables on which they depend, in fundamentally different 
ways.

An exception to the rule of a different operator treatment for first- and 
second-class constraints is offered by the {\it projection operator 
method\/} approach to the quantization of systems with constraints 
\cite{KORIG,UNIV,SCHAL}. The projection operator method was already 
discussed in P-I, and that discussion also included several elementary 
applications of the technique to simple, few degree-of-freedom systems. 
We do not repeat that discussion here. Instead, we simply observe that, 
rather than impose the self-adjoint quantum constraints in the idealized 
(Dirac) form $\Phi_\a\s|\psi\>_{phys}=0$, $\a\in\{1,\ldots,A\}$, on 
vectors $|\psi\>_{phys}$ in a putative physical Hilbert space, 
$|\psi\>_{phys}\in{\frak H}_{phys}$, we define a (possibly regularized) 
$ {\frak H}_{phys}\equiv \E \s{\frak H}$, in which $\E$ denotes a 
{\it projection operator} defined by
  \bn  \E=\E(\Sigma_\a\Phi_\a^2\le\delta(\hbar)^2)\;,  \en
where $\delta(\hbar)$ is a positive {\it regularization parameter} 
(not a $\delta$-function!)~and we have assumed that 
$\Sigma_\a\Phi_\a^2$ is self adjoint.
As a final step, the parameter $\delta(\hbar)$ is reduced as much as 
required, and, in particular, when some second-class constraints are 
involved, $\delta(\hbar)$ ultimately remains strictly positive. This 
general procedure treats all constraints simultaneously and treats them 
all on an equal basis.

\subsection{Why ultralocal fields are relevant}
In quantizing any theory with constraints, including reparametrization 
invariant theories, we invariably follow the rule: {\it Quantize first, 
reduce second}, also used by Dirac \cite{DIRAC}. $[\!\![${\bf Remark:} 
Some other quantization procedures reduce (i.e., impose constraints) 
first and quantize second, a scheme that may lead to different and 
generally incorrect results, especially when the physical phase space 
(the quotient of the constraint surface by the gauge transformations) 
is non-Euclidean \cite{SHAB}. We do not comment further on these 
alternative procedures.$]\!\!]$ It must be strongly emphasized that 
in the initial quantization phase of this pair of operations, one must 
remain neutral toward, or even better, blind to any specifics of the 
constraints to be imposed. In particular, in the initial, quantization 
phase, reparametrization invariance requires that basic fields are 
statistically independent at any spatial separation; correlations between 
fields that are spatially separated, and which originate from the 
particular physics of the theory under discussion, will arise only after 
the relevant constraints are imposed. Before the constraints are ever 
discussed, the nature of the field operators is {\it ultralocal}, the 
name given to field-operator representations which are, in fact, 
statistically independent for all distinct spatial points. The primary 
representation already presented in P-I for the basic local affine 
quantum field operators, and before constraints have been introduced, 
is for this very reason ultralocal in character. In the present paper, 
we shall discuss the relevant constraints for gravity and argue, to the 
extent possible at present, that the imposition of the constraints leads 
to field operator representations that are no longer ultralocal in 
character. Indeed, we shall argue that all traces of an initial 
ultralocal representation disappear when the constraints are fully 
enforced. $[\!\![${\bf Remark:} As an illustration of these general 
procedures, it is important at this point to observe that a relativistic 
free field of mass $m$ (and, in addition, well-defined interacting fields) 
can be quantized starting from a reparametrization invariant formulation 
and with an initial field operator representation that is ultralocal in 
nature \cite{REL}. After imposing the appropriate constraint, it may be 
shown, by a suitable modification of the reproducing kernel---see Sec.~IV 
for a brief discussion---how the conventional and {\it non\/}ultralocal 
field operator representation for the relativistic free field of an 
{\it arbitrary} mass $m$ emerges, as well as how the conventional propagator 
for such a system also arises.$]\!\!]$

\subsection{Outline of remaining sections}
In Sec.~II, we discuss in some detail the formulation of the initial 
stage of quantization in the absence of the gravitational constraints. 
In Sec.~III, the gravitational constraints are introduced, and a novel, 
but generally familiar, functional integral representation of the desired 
expressions is developed. Finally, in Sec.~IV, and aided by the use of 
several analogies, we present a lengthy discussion of the virtues, as we 
see them, of the specific functional integral representation developed in 
this article. 
\section{\sc REPRODUCING KERNEL FOR THE \\
ORIGINAL HILBERT SPACE, AND ITS\\FUNCTIONAL INTEGRAL \\REPRESENTATION}
\subsection{Basic field operator representation and the \\original Hilbert 
space}
For reasons briefly sketched above, we adopt as basic local field operators 
the positive-definite, local self-adjoint metric operators $\hg_{ab}(x)$, 
and the
local self-adjoint ``scale'' operators $\hp_d^c(x)$, which satisfy the 
affine commutation relations (\ref{afc}) in the ``original'' Hilbert space 
$\frak H$. Here $x=\{x^a\}_{a=1}^3$ denotes the coordinates (each with the 
dimension of $length$) of a point in a classical topological space 
$\cal S$, which, for the sake of discussion, we may assume is topologically 
equivalent to ${\mathbb R}^3$. The term $\delta(x,y)$ which enters the 
affine commutation relations is a Dirac $\delta$-function ``scalar density'' 
with dimensions $(length)^{-3}$. Moreover, the affine commutation relations 
uniquely tell us that $\hp_d^c(x)$ transforms as a mixed-valence tensor 
density of weight one and has the engineering dimensions of an $action$ 
(due to $\hbar$) times $(length)^{-3}$, i.e., ${\sf M/LT}$ in terms of 
$mass$ $(\sf M)$, $length$ $(\sf L)$, and $time$ $(\sf T)$. We require 
that $\hg_{ab}(x)$ transform as a covariant tensor of rank two, and 
that $\hg_{ab}(x)$ is dimensionless. 

\subsubsection*{An auxiliary---but temporary---structure}
For purposes of the present section, we shall need to augment the classical 
topological space $\cal S$ with one additional structure, namely a volume 
form. Specifically, we adopt a real, positive scalar density of weight 
one, $b(x)$, $0<b(x)<\infty$, $x\in {\cal S}$, with which we may define, 
at each point, an invariant volume form $dV\equiv b(x)\,d^3\!x$. 
Additionally, in keeping with its transformation properties, we further 
insist that $b(x)$ has the dimensions of ${\sf L}^{-3}$. 

{\it The introduction of the auxiliary structure represented by $b(x)$ 
is {\bf required} before the constraints are introduced, but whatever 
choice is made, it will {\bf disappear completely} after the constraints 
are fully enforced}. 

More explicitly, several arguments are offered below to justify the 
introduction of the auxiliary structure represented by the function 
$b(x)$, $x\in{\cal S}$, prior to the introduction of the constraints. 
In Secs.~III and IV, we will observe that we expect all the irrelevant 
freedom present in $b(x)$ to be ``squeezed out'' by the constraints and 
replaced with another, {\it non\/}arbitrary structure with the same 
dimensions. This is exactly the process observed in other cases that 
have been explicitly dealt with, and the particular case of a 
reparametrization invariant relativistic free field will be summarized 
in Sec.~IV  \cite{REL}. 

\subsubsection*{Original reproducing kernel}
Dual to the local operators $\hg_{ab}(x)$ and $\hp^c_d(x)$, we next 
introduce two real, $c$-number functions $\pi^{ab}(x)$ [$\,=\pi^{ba}(x)$] 
and $\gamma^d_c(x)$, which, initially, may be taken as smooth functions 
of compact support. Here $\pi^{ab}$ transforms as a contravariant tensor 
density of rank two and has dimensions $\sf M/LT$, while $\gamma^d_c$ 
transforms as a mixed tensor and is dimensionless. With $|\eta\>$---called 
the {\it fiducial vector}---an as yet unspecified unit vector in $\frak H$, 
we consider the set of unit vectors (in units where $\hbar=1$) each of 
which is given by
\bn
  |\pi,\gamma\>\equiv e^{i\tint \pi^{ab}(x)\,\hg_{ab}(x)\,d^3\!x}\,
e^{-i\tint\gamma^d_c(x)\,{\hat\pi}^c_d(x)\,d^3\!x}\,|\eta\>\;.  \en
As $\pi^{ab}$ and $\gamma^d_c$ range over the space of smooth functions of 
compact support, such vectors form a set of {\it coherent states}. 

The complex functional $\<\pi'',\gamma''|\pi',\gamma'\>$ formed by the 
inner product of two such coherent states will be a functional of 
fundamental importance in the present study of gravity. In particular, 
the functional $\<\pi'',\gamma''|\pi',\gamma'\>$, whatever form it takes, 
is manifestly a positive-definite functional that fulfills the defining 
condition that
 \bn
   \sum_{j,k=1}^J\a^*_j\,\a_k\,\<\pi_j,\gamma_j|\pi_k,\gamma_k\>\ge0  \en
for general sets $\{\a_j\}$ and $\{\pi_j,\gamma_j\}$ for any $J<\infty$. 
Furthermore, $\<\pi'',\gamma''|\pi',\gamma'\>$ is always a continuous 
functional in some natural functional topology, e.g., a topology defined 
by the particular expression itself \cite{HEG}. As a continuous, 
positive-definite functional, it follows that we may adopt the expression 
$\<\pi'',\gamma''|\pi',\gamma'\>$ as a {\it reproducing kernel}, and use 
it to define an associated {\it reproducing kernel Hilbert space} 
$\cal C$ \cite{MESH}. Let two elements of a dense set of elements in 
$\cal C$ be given by
  \bn  && \psi(\pi,\gamma)\equiv \sum_{j=1}^J\a_j 
\<\pi,\gamma|\pi_j,\gamma_j\>\;,\hskip.6cm J<\infty\;, \no\\
  && \phi(\pi,\gamma)\equiv \sum_{k=1}^K\beta_k \<\pi,
\gamma|{\o\pi}_k,{\o\gamma}_k\>\;,\hskip.6cm K<\infty\;,  \en
where $\{{\o\pi}_k,{\o\gamma}_k\}_{k=1}^K$ denotes another independent 
set of (real) fields.
These are continuous functionals of the fields $\pi$ and $\gamma$. As 
the inner product of these two elements we adopt
\bn  (\psi,\phi)\equiv\sum_{j=1}^J\sum_{k=1}^K\a^*_j\,\beta_k 
\<\pi_j,\gamma_j|{\o\pi}_k,{\o\gamma}_k\>\;.  \en
We complete the space of functions by including the limit point of all 
Cauchy sequences in the norm $\|\psi\|\equiv (\psi,\psi)^{1/2}$.

The result of the above construction is the (separable) reproducing kernel 
Hilbert space $\cal C$ composed of bounded, continuous functionals. 
Moreover, the Hilbert space $\cal C$ provides an especially useful 
functional representation of our original Hilbert space, which was 
referred to as the ``primary container'' in P-I.  

\subsection{Choice of the fiducial vector and explicit form of the
reproducing kernel}
As argued in Sec.~I, the representation of the basic field operators 
$\hg_{ab}$ and $\hp^c_d$ must be ultralocal prior to the introduction of 
any constraints. To fulfill this requirement, it is necessary that
\bn
  \<\pi'',\gamma''|\pi',\gamma'\>=\exp\{-\tint b(x)\,d^3\!x\,L[\pi''(x),
\gamma''(x);\pi'(x),\gamma'(x)]\}  \en
for some dimensionless scalar function $L$. This function is determined 
by the representation of the affine field operators and the fiducial 
vector $|\eta\>$, and as minimum conditions we require that
\bn
  && \<\eta|\s\hg_{ab}(x)\s|\eta\>\equiv{\tilde g}_{ab}(x)\;,\\
  && \hskip.08cm\<\eta|\s\hp^c_d(x)\s|\eta\>\equiv 0\;.  \en
Here, ${\tilde g}(x)\equiv\{{\tilde g}_{ab}(x)\}$ is a fixed, smooth, 
positive-definite metric function determined by the choice of $|\eta\>$ 
(see \cite{P-I}). Whether $\cal S$ is compact or noncompact, the choice of 
${\tilde g}(x)$ will determine the topology of the space-like surfaces under 
consideration; if $\cal S$ is noncompact, then ${\tilde g}(x)$ also 
determines the asymptotic form of the space-like surfaces under consideration.

For reasons to be offered below, we choose $|\eta\>$ so that the overlap 
function of two coherent states is
given (when $\hbar=1$) by 
 \bn &&\hskip0cm\<\pi'',\gamma''|\pi',\gamma'\> \no\\
  &&\hskip.3cm =\exp\bigg[-2\int b(x)\,d^3\!x\,\no\\
&&\hskip.1cm\times\ln\bigg(\frac{\det\{\half[g''^{ab}(x)+g'^{ab}(x)]+
\half ib(x)^{-1}[\pi''^{ab}(x)-\pi'^{ab}(x)]\}}{\{\det[g''^{ab}(x)]\,
\det[g'^{ab}(x)]\}^{1/2}}\bigg)\bigg]\;. \label{e18}\en
Several comments about this basic expression are in order. 

Initially, regarding (\ref{e18}), we observe that $\gamma''$ and 
$\gamma'$ do {\it not} appear in the explicit functional form given. 
In particular, the smooth matrix $\gamma$ has been replaced by the smooth 
matrix $g$ which is defined at every point by
 \bn  g(x)\equiv e^{\gamma(x)/2}\,{\tilde g}(x)\,e^{\gamma(x)^T/2}
\equiv\{g_{ab}(x)\}\;, \en
where $\gamma(x)^T$ denotes the transpose of the matrix $\gamma(x)$.
Observe that the so-defined matrix $\{g_{ab}(x)\}$ is manifestly positive 
definite for all $x$. The map $\gamma\ra g$ is clearly many-to-one since 
$\gamma$
 has {\it nine} independent variables at each point while $g$, which is 
symmetric, has only {\it six}. 
In view of this functional dependence {\it we may denote the given 
functional in (\ref{e18}) by $\<\pi'',g''|\pi',g'\>$, and henceforth 
we shall adopt this notation exclusively}. 

\subsubsection*{Single affine matrix degree of freedom}
An elementary example of the notational change from $\gamma$ to $g$ may 
be seen quite directly in what occurs for a {\it single} affine matrix 
degree of freedom (in contrast to a field of such degrees of freedom). To 
that end,
and following Ref.~11 closely, we introduce the Lie algebra for affine 
matrix self-adjoint operator degrees of freedom composed of the symmetric 
$3\times3$ matrix $\{\sigma_{ab}\}$ and the $3\times3$ matrix 
$\{\kappa^c_d\}$, which together obey the affine commutation relations 
[cf., (\ref{afc})]
  \bn  &&\hskip.25cm [\kappa^a_b,\kappa^c_d]=i\s\half\s(\delta^c_b\s
\kappa^a_d-\delta^a_d\s\kappa^c_b)\;, 
\no\\
  &&\hskip.125cm[\sigma_{ab},\kappa^c_d]=i\s\half\s(\delta^c_a\s
\sigma_{db}+\delta^c_b\s\sigma_{ad})\;,  \\
  &&[\sigma_{ab},\sigma_{cd}]=0\;. \no \en
We choose the faithful, irreducible representation for which the operator 
matrix $\{\sigma_{ab}\}$ is symmetric and positive definite, and which is 
unique up to unitary equivalence. Furthermore, we choose a representation 
which diagonalizes $\{\sigma_{ab}\}$ as $k\equiv\{k_{ab}\}$, which we 
refer to as the $k$-representation. In the associated $L^2$ representation 
space, and for arbitrary real matrices $F=\{F^{ab}\}$, $F^{ba}=F^{ab}$, 
and $B=\{B^c_d\}$, it follows that
 \bn && U[F,B]\s\psi(k)\equiv e^{iF^{ab}\sigma_{ab}}\,
e^{-iB^c_d\kappa^d_c}\s\psi(k)\no\\
    &&\hskip2.33cm= (\det[S])^{2}\,e^{iF^{ab}k_{ab}}\s\psi(SkS^T)\;,  \en
where $S\equiv e^{-B/2}=\{S^a_b\}$ and $(SkS^T)_{ab}\equiv 
S_a^ck_{cd}S^d_b$. The given transformation is unitary within the inner 
product defined by
 \bn \int_+\psi(k)^*\s\psi(k)\,dk \;, \en
where $dk\equiv\Pi_{a\le b}\s dk_{ab}$, and the ``$+$'' sign denotes an 
integration over only that part of the six-dimensional $k$-space where 
the elements form a symmetric, positive-definite matrix, $\{k_{ab}\}>0$. 
To define coherent states we choose an extremal weight vector,
  \bn  \eta(k)\equiv C\s(\det[k])^{\beta-1}\,
e^{-\beta\s\tr[{\tilde G}^{-1}k]}\;,  \en
where $\beta>0$, ${\tilde G}=\{{\tilde G}_{ab}\}$ is a fixed 
positive-definite matrix, $C$ is determined by normalization, and 
tr denotes the trace. This choice leads to the expectation values
 \bn  && \<\eta|\s\sigma_{ab}\s|\eta\>=\int_+\eta(k)^*\s k_{ab}\s
\eta(k)\,dk={\tilde G}_{ab}\;,  \\
    &&\hskip.3cm\<\eta|\s\kappa^c_d\s|\eta\>=\int_+\eta(k)^*\s 
\kappa^c_d\s\eta(k)\,dk=0\;.  \en
In the $k$-representation, it follows that the affine matrix 
coherent states are given by 
  \bn  \<k|F,B\>\equiv C\s(\det[S])^{2}\s
(\det[SkS^T])^{\beta-1}\,e^{i\s\tr[Fk]}\,
e^{-\beta\s\tr[{\tilde G}^{-1}SkS^T]} \;.  \en
Observe that what really enters the functional argument is the 
positive-definite matrix $G^{-1}\equiv S^T{\tilde G}^{-1}\s S$ where 
we set $G\equiv\{G_{ab}\}$. Thus without loss of generality we can drop 
the label $B$ (or equivalently $S$) and replace it with $G$. Hence the 
affine matrix coherent states become
  \bn  \<k|F,G\>\equiv C'\s(\det[G^{-1}])^{\beta}\s
(\det[k])^{\beta-1}\,e^{i\s\tr[Fk]}\,e^{-\beta\s\tr[G^{-1}k]}\;,  \en
where $C'=C\s(\det[{\tilde G}])^\beta$ is a new normalization constant.
It is now straightforward to determine that
  \bn   &&\hskip-1.5cm\<F'',G''|F',G'\>=\int_+\<F'',G''|k\>\s
\<k|F',G'\>\,dk \no\\
  &&  \hskip1.31cm=\bigg[\frac{\{\det[G''^{-1}]\det[G'^{-1}]\}^{1/2}}
{\det\{\half[(G''^{-1}+G'^{-1})+i\beta^{-1}(F''-F')]\}}\bigg]^{2\beta} \;.  
\label{h30}\en
In arriving at this result, we have used normalization of the coherent 
states to eliminate the constant $C'$. 

\subsubsection*{Lattice construction}
Suppose now that we consider an independent lattice of such matrix degrees 
of freedom and build the corresponding coherent state overlap as the product 
of ones just like (\ref{h30}). Let $\bf n$ label a lattice site and let 
${\bf n}\in{\bf N}$, which in turn is a finite subset of ${\mathbb Z}^3$. 
In that case the coherent state overlap is given by
 \bn &&\<F'',G''|F',G'\>_{\bf N} \no\\
&&\hskip1cm=\prod_{{\bf n}\in{\bf N}}
\bigg[\frac{\{\det[G''^{-1}_{[{\bf n}]}]
\det[G'^{-1}_{[{\bf n}]}]\}^{1/2}}
{\det\{\half[(G''^{-1}_{[{\bf n}]}+
G'^{-1}_{[{\bf n}]})+i\beta^{-1}_{[{\bf n}]}(F''_{[{\bf n}]}-
F'_{[{\bf n}]})]\}}\bigg]^{2\beta_{[{\bf n}]}} \;.  \label{i24}\en
As our next step we wish to take a limit in which the number of 
independent matrix degrees of freedom tends to infinity in such a way 
that not only does the lattice size diverge but also the lattice spacing 
tends to zero so that, loosely speaking, the lattice points approach the 
points of the space $\cal S$. In order for the limit to be nonzero, it is 
necessary that the exponent $\beta_{[{\bf n}]}\ra0$ in a suitable way. To 
that end we set
 \bn  \beta_{[{\bf n}]}\equiv b_{[{\bf n}]}\s\De  \;,  \en
where $\De$ has the dimensions ${\sf L}^3$, and thus $b_{[{\bf n}]}$ has 
the dimensions ${\sf L}^{-3}$. In addition we need to let $F^{ab}_{[{\bf n}]}
\equiv \pi^{ab}_{[{\bf n}]}\s\De$, and we rename $G_{ab\s{[{\bf n}]}}$ as 
$g_{ab\s{[{\bf n}]}}$ and call the matrix elements of $G^{-1}_{[{\bf n}]}$ 
by $g^{ab}_{[{\bf n}]}$. With these changes 
(\ref{i24}) becomes 
\bn &&\<\pi'',g''|\pi',g'\>_{\bf N} \no\\
&&\hskip1cm\equiv\prod_{{\bf n}\in{\bf N}}\bigg[
\frac{\{\det[g''^{ab}_{[{\bf n}]}]\det[g'^{ab}_{[{\bf n}]}]\}^{1/2}}
{\det\{\half[(g''^{ab}_{[{\bf n}]}+g'^{ab}_{[{\bf n}]})+i\s 
b^{-1}_{[{\bf n}]}(\pi''^{ab}_{[{\bf n}]}-\pi'^{ab}_{[{\bf n}]})]\}}
\bigg]^{2\s b_{[{\bf n}]}\s\De} \;.  \label{i25}\en
Finally, we take the limit as described above and the result is given by 
(\ref{e18}), namely,
 \bn  &&\hskip-.8cm\<\pi'',g''|\pi',g'\> \no\\
 &&\hskip-.3cm =\exp\bigg[-2\int b(x)\,d^3\!x\,\no\\
&&\hskip.1cm\times\ln\bigg(\frac{\det\{\half[g''^{ab}(x)+g'^{ab}(x)]+
\half ib(x)^{-1}[\pi''^{ab}(x)-\pi'^{ab}(x)]\}}{\{\det[g''^{ab}(x)]\,
\det[g'^{ab}(x)]\}^{1/2}}\bigg)\bigg]\;.  \label{i22}\en

In this way we see how the continuum result may be obtained as a limit 
starting from a collection of independent affine matrix degrees of freedom. 
The necessity of ending with an integral over the space $\cal S$ has 
directly led to the requirement that we introduce the scalar density $b(x)$.

\subsection{Additional arguments favoring $b(x)$}
As a further general comment about (\ref{e18}) or (\ref{i22}) we observe 
that $\<\pi'',g''|\pi',g'\>$ is {\it invariant} under general (smooth, 
invertible) coordinate transformations $x\ra {\o x}={\o x}(x)$, and we say 
that the given expression characterizes a {\it diffeomorphism invariant 
realization} of the affine field operators. This property holds, in part, 
because $b(x)$ transforms as a scalar density in both places that it appears. 
Thus $b(x)$, which has the dimensions of ${\sf L}^{-3}$, plays an essential 
dimensional and transformational role in each place that it appears. 
Note that if $\hbar$ is explicitly introduced into (\ref{i22}) or 
(\ref{e18}), it appears only in the change $[\pi''^{ab}(x)-\pi'^{ab}(x)]
\ra[\pi''^{ab}(x)-\pi'^{ab}(x)]/\hbar$.  If one insisted on building an 
acceptable ultralocal positive-definite functional using only 
$\pi''^{ab}(x)$, $g''^{ab}(x)$, $\pi'^{ab}(x)$, $g'^{ab}(x)$, and 
$\hbar$, then that construction would not appear to be possible simply 
on dimensional grounds. 

As another argument for the appearance of $b(x)$, we observe that 
(\ref{i22}) involves not only the $c$-number fields $\pi$ and $g$ 
(or $\gamma$), but the particular representation of the local operators 
$\hg_{ab}$ and $\hp^c_d$ as well as the choice of the fiducial vector 
$|\eta\>$. It is entirely natural that the function $b(x)$ may emerge 
as a needed functional parameter in defining the operator representation 
and/or the vector $|\eta\>$, and this property is explicitly illustrated 
in P-I. 

As a final argument for the appearance of the scalar density $b(x)$, we 
briefly recall properties of the local operator product for ultralocal 
affine field operators \cite{P-I}. In particular, the formal local product 
reads
 \bn  \hg_{ab}(x)\s\hg_{cd}(x)=\delta(x,x)\s{\hat E}_{abcd}(x)+l.s.t.\;. \en
Here ${\hat E}_{abcd}(x)$ is a local fourth-order covariant tensor density 
operator of weight $-1$, $\delta(x,x)$ is a divergent multiplier with 
dimensions ${\sf L}^{-3}$, which arises when the ``scalar density'' delta 
function $\delta(x,y)$ is evaluated at coincident points, and $l.s.t.$ 
denotes ``less singular terms''. Before adopting the proper local operator 
product, we introduce a scalar density $b(x)$, $0< b(x)<\infty$, with 
dimensions ${\sf L}^{-3}$, and consider
 \bn  \hg_{ab}(x)\s\hg_{cd}(x)= b(x)\s[\s b(x)^{-1}\,\delta(x,x)\s]\s
{\hat E}_{abcd}(x)+l.s.t.\;. \en
Finally, we choose
 \bn [\hg_{ab}(x)\s\hg_{cd}(x)]_R\equiv b(x)\s{\hat E}_{abcd}(x)\;\;
(=\hg_{ab}(x)\s\hg_{cd}(x)/[\s b(x)^{-1}\delta(x,x)\s]) \en
as the proper renormalized (subscript R) local operator product. 
(Limits involving test function sequences offer a mathematically precise 
construction.) The given choice leads to a local operator that transforms 
as a tensor in the natural fashion and, moreover, carries the natural 
engineering dimensions. To achieve this desirable property in a local 
product {\it requires} the introduction of an auxiliary scalar density 
$b(x)$.

For additional properties regarding local products of the relevant affine 
field operators, see P-I. Essentially, all these properties are almost 
entirely based on the analysis of local operator products in scalar 
ultralocal field theories, an analysis which is described in detail, for 
example, in \cite{KBOOK}. 

We have offered several reasons for the appearance of the scalar density 
$b(x)$ at the present stage of the analysis. However, we emphasize once 
again that $b(x)$ will disappear when the constraints are fully enforced, 
whatever choice was originally made. An example of the process by which 
this fundamental transformation takes place is presented in Sec.~IV. 

\subsection{Functional integral representation for the \\coherent state 
overlap functional}
For further analysis, especially when we take up the issue of introducing 
the constraints in Sec.~III, it is useful to introduce an alternative 
representation of the functional expression (\ref{i22}). The alternative 
representation we have in mind is that of a specific functional integral, 
which, indeed, has already been introduced in P-I. That such a 
representation should exist is an immediate consequence of the fact that 
(\ref{i22}) fulfills a complex polarization condition, which then leads to 
(\ref{i22}) being annihilated by  a negative, second-order functional 
derivative operator. Exponentiating this operator, times a parameter 
$\nu>0$, letting the resultant operator act on general functionals of 
$\pi''$ and $g''$, and including any necessary $\nu$-dependent prefactor, 
will, in the limit $\nu\ra\infty$, lead to a dense set of functions in the 
reproducing kernel Hilbert space $\cal C$. Alternatively, letting the same 
operator act on a suitable $\delta$-functional will lead to the expression 
(\ref{i22}). The functional integral of interest arises in this last 
expression by introducing the analog of a Feynman-Kac-Stratonovich 
representation. The mathematics behind these foregoing several sentences 
is well illustrated in P-I for both a simple, single affine degree-of-freedom 
example as well as for the affine field theory.

The result of the operations outlined above leads to a functional integral 
representation for $\<\pi'',g''|\pi',g'\>$ in (\ref{i22}), which is given 
(for $\hbar=1$) by
  \bn  &&\<\pi'',g''|\pi',g'\>\no\\
  &&\hskip.8cm=\exp\bigg[-2\int b(x)\,d^3\!x\,\no\\
&&\hskip1.4cm\times\ln\bigg(\frac{\det\{\half[g''^{ab}(x)+g'^{ab}(x)]+
\half ib(x)^{-1}[\pi''^{ab}(x)-\pi'^{ab}(x)]\}}{\{\det[g''^{ab}(x)]\,
\det[g'^{ab}(x)]\}^{1/2}}\bigg)\bigg] \no\\
&&\hskip.8cm=\lim_{\nu\ra\infty}\,{\o{\cal N}}_{\nu}\,\int 
\exp[-i\tint g_{ab}\s{\dot\pi}^{ab}\,d^3\!x\,dt]\no\\
  &&\hskip1.4cm\times\exp\{-(1/2\nu)\tint[b(x)^{-1}g_{ab}g_{cd}
{\dot\pi}^{bc}{\dot\pi}^{da}+b(x)g^{ab}g^{cd}{\dot g}_{bc}{\dot g}_{da}]
\,d^3\!x\,dt\}\no\\
&&\hskip2.3cm\times\Pi_{x,t}\,\Pi_{a\le b}\,d\pi^{ab}(x,t)\,dg_{ab}(x,t) 
\label{e20}\;.  \en
Here, because of the way the new independent variable $t$ appears on the 
right-hand side of this expression, it is natural to interpret $t$, 
$0\le t\le T$, $T>0$ as coordinate ``time''. The fields on the right-hand 
side all depend on space and time, i.e., $g_{ab}=g_{ab}(x,t)$, 
${\dot g}_{ab}=\d g_{ab}(x,t)/\d t$, etc., and, importantly, the 
integration domain of the formal measure is strictly limited to the 
domain where $\{g_{ab}(x,t)\}$ is a positive-definite matrix for all $x$ 
and $t$. For the boundary conditions, we have $\pi'^{ab}(x)
\equiv\pi^{ab}(x,0)$, $g'_{ab}(x)\equiv g_{ab}(x,0)$, as well as 
$\pi''^{ab}(x)\equiv\pi^{ab}(x,T)$,
$g''_{ab}(x)\equiv g_{ab}(x,T)$ for all $x$. Observe that the right-hand 
side holds for {\it any} $T$, $0<T<\infty$, while the middle term is 
{\it independent of $T$ altogether}. 

Although the functional integral on the right-hand side is formal it 
nevertheless conveys a great deal of information. Let us first examine 
it from a dimensional and transformational standpoint. As presented, 
$\hbar=1$; to see where $\hbar$ would appear we may simply replace each 
$\pi^{ab}$ by $\pi^{ab}/\hbar$. With $\nu$ having the dimensions of 
${\sf T}^{-1}$ and ${\o{\cal N}}_\nu$ absorbing any remaining dimensions 
from the formal measure, then the right-hand side of (\ref{e20}) is 
dimensionally satisfactory. From the point of view of (formal) 
transformations under coordinate changes, it is clear, with 
${\o{\cal N}}_{\nu}$ transforming appropriately, that the right-hand 
side is formally invariant under coordinate transformations involving 
the spatial coordinates alone. $(\!\!(${\bf Remark:} A discussion about 
transformations of the right-hand side under spatially dependent 
transformations of the time coordinate has been given in P-I and is not 
repeated here. It is clear that the result of the limit $\nu\ra\infty$ 
on the right-hand side must be invariant under all such transformations 
simply because the middle term is independent of the time variable 
altogether$!)\!\!)$ 

As presented---and indeed as originally derived---the {\it result} of the 
functional integral (the middle term) came before the functional integral 
{\it representation} of that result (the right-hand side). However, we can 
also interpret (\ref{e20}) in the opposite order, that is, to presume 
that the functional integral (right-hand side) is primary and that the 
answer (the middle term) is the result of evaluating the functional 
integral. This perspective encourages us to examine the expression in the 
integrand of the functional integral somewhat more carefully from a 
traditional standpoint. We first observe that the formal, flat part of the 
measure has the expected appearance of the {\it canonical measure} for a 
conventional, canonical functional integral quantization of gravity. The 
phase factor contains an acceptable classical symplectic potential term in 
a formal functional integral for gravity in which the rest of the classical 
action---the terms involving the constraints---are absent. 
$(\!\!(${\bf Remark:} This characterization is, of course, quite appropriate 
since we still are in the first phase of our dual approach: quantize first, 
reduce second. The patient reader will be rewarded with the addition of the 
expected constraints and Lagrange multiplier terms in the next 
section.$)\!\!)$ The second, $\nu$-dependent factor in the integrand 
serves as a {\it regularizing factor} for the functional integral. 
Formally, as $\nu\ra\infty$, such a factor disappears from the integrand 
leaving the expected (pre-constraint) formal functional integral integrand, 
such as it is. However, the $\nu$-dependent term plays a fundamentally 
important role within the integral itself since it {\it literally serves 
to define the functional integral}. 

It is important to make this last point quite clear, and for that purpose 
we temporarily discuss the formal expression (with $\hbar=1$ again)
  \bn && d\mu^\nu(\pi,g)\no\\
&&\hskip.4cm={\cal M}_\nu\,\exp\{-(1/2\nu)\tint[b(x)^{-1}g_{ab}g_{cd}
{\dot\pi}^{bc}{\dot\pi}^{da}+
b(x)g^{ab}g^{cd}{\dot g}_{bc}{\dot g}_{da}]\,d^3\!x\,dt\}\no\\
&& \hskip1.5cm\times\Pi_{x,t}\,\Pi_{a\le b}\,d\pi^{ab}(x,t)\,
dg_{ab}(x,t)\;.  \label{f17}\en
We assert that for fixed $b(x)$, $0<b(x)<\infty$ and fixed $\nu$, 
$0<\nu<\infty$, this expression characterizes a {\it bona fide, 
countably-additive, positive measure}, $\mu^\nu$, on the space of 
generalized functions $\pi^{ab}=\pi^{ab}(x,t)$ and $g_{ab}=g_{ab}(x,t)$, 
where, for any nonvanishing $u^a$ and any nonvanishing, nonnegative test 
function $f(x)\ge0$, the positive-definite matrix condition $\tint f(x)\s 
u^a\s g_{ab}(x,t)\s u^b\,d^3\!x>0$ holds for (almost) all $t$, $0<t<T$. The 
fields $\pi^{ab}$ and $g_{ab}$ satisfy the boundary conditions at $t=0$ and 
$t=T$ given previously. The factor ${\cal M}_\nu$ is adjusted so that the 
measures $\mu^\nu$ form a semi-group with respect to combining time 
intervals, e.g., $0\ra T$, $T>0$, and then $T\ra T+T'$, $T'>0$, being 
equivalent to $0\ra T+T'$. If $\{h_p(x)\}_{p=1}^\infty$ denotes an 
orthonormal set of test functions defined so that
  \bn  && \tint h_p(x)\s h_q(x)\s b(x)\,d^3\!x=\delta_{pq}\;,  \\
  && \hskip.07cm b(x)\s\Sigma_{p=1}^\infty h_p(x)\s h_p(y)=\delta(x,y)\;,  \en
then we assert that there exist finite, nonzero constants $N^\nu_P$ for 
all $\nu$ and all $P\in\{1,2,3,\ldots\}$ such that
  \bn N^\nu_P\int\exp[\s-i\s\Sigma_{p=1}^P\tint g_{ab(p)}(t)\s
{\dot\pi}^{ab}_{(p)}(t)\,dt\s]\,d\mu^\nu(\pi,g) \en
is well defined. In this expression,
  \bn && g_{ab(p)}(t)\equiv\tint h_p(x)\s g_{ab}(x,t)\s b(x)\,d^3\!x\;,  \\
      && \hskip.25cm{\dot\pi}^{ab}_{(p)}(t)\equiv \tint h_p(x)\s
{\dot\pi}^{ab}(x,t)\,d^3\!x\;.  \en
Moreover, the set of constants $\{N^\nu_P\}$ may be chosen so that
 \bn &&\hskip-.5cm\<\pi'',g''|\pi',g'\>\no\\
   &&\hskip0cm\equiv\lim_{P\ra\infty}\,\lim_{\nu\ra\infty}\,N^\nu_P\,
\int\exp[\s-i\s\Sigma_{p=1}^P\tint g_{ab(p)}(t)\s{\dot\pi}^{ab}_{(p)}(t)\,
dt\s]\,d\mu^\nu(\pi,g) \;.\en
This is one of the ways that the formal functional integral (\ref{e20}) 
can be given a rigorous meaning.

There is another way to give rigorous meaning to (\ref{e20}) that we 
would also like to discuss. In this procedure we use a spatial lattice 
but keep the time variable $t$ continuous. This regularization scheme 
takes us back to the lattice construction given earlier [cf., (\ref{i25})], 
except now we add a phase-space path integral representation as well. For 
present purposes we again introduce the symbol $\De$, with dimensions 
${\sf L}^{3}$, to denote a uniformly (coordinate) sized, small spatial 
cell. Then the lattice regularized path integral expression for 
$\<\pi'',g''|\pi',g'\>_{\bf N}$ in (\ref{i25}) is given by \cite{GUTZ}
 \bn
   && {\o{\cal N}}^{\bf N}_\nu\,\int e^{-i\Sigma_{\bf n}
\tint g_{ab[{\bf n}]}\s{\dot\pi}^{ab}_{[{\bf n}]}\,\De\,dt}\no\\
&&\hskip.5cm\times \exp\{-\halfnu\Sigma_{\bf n}\tint[b_{[{\bf n}]}^{-1}
g_{ab[{\bf n}]}g_{cd[{\bf n}]}{\dot\pi}^{bc}_{[{\bf n}]}
{\dot\pi}^{da}_{[{\bf n}]}+b_{[{\bf n}]}g^{ab}_{[{\bf n}]}
g^{cd}_{[{\bf n}]}{\dot g}_{bc[{\bf n}]}{\dot g}_{da[{\bf n}]}]
\,\De\,dt\}\no\\
  &&\hskip2cm \times\Pi_{{\bf n},t}\,\Pi_{a\le b}\,
d\pi^{ab}_{[{\bf n}]}(t)\,dg_{ab[{\bf n}]}(t) 
\;, \label{g24}\en
where ${\bf n}\in{\bf N}$, which itself is a finite subset of 
${\mathbb Z}^3$. Here $\pi^{ab}_{[{\bf n}]}$, $g_{ab[{\bf n}]}$, 
and $b_{[{\bf n}]}$ represent average field values in the cell 
${\bf n}$, and $b_{[{\bf n}]}^{-1}=1/b_{[{\bf n}]}$. Next, we observe 
that there is a countably-additive, pinned, Brownian-motion measure 
formally defined by
 \bn && \hskip1cm d\mu^\nu_{\bf N}(\pi,g)\no\\
   &&\hskip-.6cm \equiv{{\cal N}}^{\bf N}_\nu\,
\exp\{-\halfnu\Sigma_{\bf n}\tint[b_{[{\bf n}]}^{-1}g_{ab[{\bf n}]}
g_{cd[{\bf n}]}{\dot\pi}^{bc}_{[{\bf n}]}{\dot\pi}^{da}_{[{\bf n}]}+
b_{[{\bf n}]}g^{ab}_{[{\bf n}]}g^{cd}_{[{\bf n}]}{\dot g}_{bc[{\bf n}]}
{\dot g}_{da[{\bf n}]}]\,\De\,dt\}\no\\
  &&\hskip2cm \times\Pi_{{\bf n},t}\,\Pi_{a\le b}\,d\pi^{ab}_{[{\bf n}]}(t)
\,dg_{ab[{\bf n}]}(t) \label{g25}\en
so that (\ref{g24}) becomes
 \bn {\o N}^{\bf N}_\nu\,\int e^{-i\Sigma_{\bf n}\tint g_{ab[{\bf n}]}\s
{\dot\pi}^{ab}_{[{\bf n}]}\,\De\,dt}\,d\mu^\nu_{\bf N}(\pi,g) \;, \en
where $\{{\o N}^{\bf N}_\nu\}$ is a set of finite constants. Moreover, these
constants may be chosen so that
 \bn \<\pi'',g''|\pi',g'\> 
 =\lim_{{\bf N}\ra\infty}\,\lim_{\nu\ra\infty}\,{\o N}^{\bf N}_\nu\,
\int e^{-i\Sigma_{\bf n}\tint g_{ab[{\bf n}]}\s{\dot\pi}^{ab}_{[{\bf n}]}\,
\De\,dt}\,d\mu^\nu_{\bf N}(\pi,g) \;. \en
Here, in the last step, the limit ${\bf N}\ra\infty$ means that $\De\ra0$ 
and ${\bf N}\ra{\mathbb Z}^3$ in such a way that all points $x\in{\cal S}$ 
are reached in a natural way. Observe that the presence of the 
continuous-time regularization factor in the formal functional integral 
for the entire space has controlled the spatial lattice regularization in 
a clear and natural fashion; although possible to introduce, no temporal 
lattice has been required to obtain a well-defined expression.

The present usage of well-defined, phase-space measures to define 
functionals integrals in the limit that the
``diffusion constant'' parameter $(\nu)$ diverges is part of the general 
program of {\it continuous-time regularization} \cite{DAUB}. It is 
noteworthy that the use of a suitable {\it phase-space metric} to control 
the Brownian motion paths invariably leads to a coherent-state 
representation for the resultant quantum amplitude. These remarks 
conclude our brief excursion into a rigorous discussion of the functional 
integrals of present interest.

\subsubsection*{Another prospective regularization}
Let us return to the formal functional integral (\ref{e20}) and examine 
that expression with regard to the scalar density $b(x)$. Superficially, 
in the limit $\nu\ra\infty$ in which the regularizing term proportional to 
$(1/2\nu)$ in the exponent of the integrand formally vanishes, one might 
naively expect that the result of the integral would be independent of the 
function $b(x)$. But, no, that naive expectation is false since the result 
of that integral and subsequent limit, i.e., the middle term of (\ref{e20}), 
evidently depends importantly on $b(x)$. At first glance, that dependence 
seems highly unnatural. It may be thought---as the author did for a number 
of years \cite{CARG}---that an alternative regularization expression may 
be more ``natural'' and would therefore be preferable. In particular, it 
was thought that the expression
 \bn \exp\{-(1/2\nu)\tint[g^{-1/2}g_{ab}g_{cd}{\dot\pi}^{bc}
{\dot\pi}^{da}+g^{1/2}g^{ab}g^{cd}{\dot g}_{bc}{\dot g}_{da}]\,
d^3\!x\,dt\}\;, \label{e27}\en
where $g=\det[g_{ab}]$, and which involves a different phase-space 
metric,
was ``better''. However, in light of the discussion in the present paper, 
it is now evident that this expression is not even dimensionally consistent! 
This defect could be rectified by the introduction of a positive constant, 
which we may call ${\tilde b}$, that stands in the place of the present $b$ 
(next to $g^{1/2}$) and carries the dimensions ${\sf L}^{-3}$. Indeed, we 
could also introduce in place of $\tilde b$ a positive scalar function 
${\tilde b}(x)$ with dimensions ${\sf L}^{-3}$. In any case, the failure 
of (\ref{e27}) purely on dimensional grounds, and the necessity thereby of 
introducing some sort of auxiliary dimensioned parameter (or function) 
surely renders (\ref{e27}) far less ``natural'' than had been previously 
assumed.

As another possible argument against (\ref{e27}), we note that if that 
form of a proposed regularization was used in an expression like (\ref{e20}) 
in place of the present form of regularization, there is absolutely no 
guarantee that the result will describe a reproducing kernel with other 
than a {\it one-dimensional reproducing kernel Hilbert space}, which is the 
general result for a ``random'' choice of phase-space metric. The fact that 
the present form of (\ref{e20}) generates a suitable, infinite-dimensional 
reproducing kernel Hilbert space is a fundamentally important feature, 
which, in our case, is a consequence of having started with an appropriate 
reproducing kernel to begin with. 

Although the author has not foreclosed any possible interest in a 
${\tilde b}$ modified version of the regularization (\ref{e27}), all 
present indications point to the version (\ref{f17}) that features the 
scalar density $b(x)$. This shift of allegiance has also been bolstered by 
the realization that the role of $b(x)$ is confined to the initial phase of 
quantization and that $b(x)$ will disappear entirely after the constraints 
are fully enforced. On the strength of this argument, it is the version 
based on ({\ref{f17}) and not (\ref{e27}) that is analyzed in the remainder 
of the present paper.

\section{\sc INTRODUCTION OF THE\\ GRAVITATIONAL CONSTRAINTS}
\subsection{Key principles in heuristic form}
There are four gravitational constraint functions, the three diffeomorphism 
constraints 
  \bn  H_a(x)=-2\pi^b_{a|b}(x)\;,  \en
where ``$_|$'' denotes covariant differentiation with respect to the spatial 
metric $g_{ab}$, and the Hamiltonian constraint, which, in suitable units 
(i.e., $c^3/G=16\pi$), reads
  \bn  H(x)= g(x)^{-1/2}[\,\pi^a_b(x)\s\pi^b_a(x)-\half\pi^a_a(x)\s
\pi^b_b(x)\,]+g(x)^{1/2}\s^{(3)}\!R(x)\;, 
\label{h45} \en
where $g(x)\equiv\det[g_{ab}(x)]$ and $^{(3)}\!R(x)$ denotes the scalar 
curvature derived from the spatial metric \cite{MTW}. Classically, these 
constraint functions vanish, and the region in phase space on which they 
vanish is called the {\it constraint hypersurface}. 

It is instructive to evaluate the classical Poisson brackets between the 
constraint fields. For this purpose, we enlist only the basic nonvanishing 
Poisson bracket given by 
 \bn  \{g_{ab}(x),\pi^{cd}(y)\}=\half(\delta^c_a\delta^d_b+\delta^c_b
\delta^d_a)\,\delta(x,y)\;. \en
It follows that
  \bn && \{H_a(x),H_b(y)\}=\delta_{,a}(x,y)\s H_b(x)-\delta_{,b}(x,y)\s 
H_a(x)\;, \label{e31}\\
    &&\hskip.15cm\{H_a(x),H(y)\}=\delta_{,a}(x,y)\s H(x)\;,  \label{e32}\\
   &&\hskip.3cm\{H(x),H(y)\}=\delta_{,a}(x,y)\s g^{ab}(x)\s H_b(x)\;.  
\label{e33}\en
In these expressions, $\delta_{,a}(x,y)\equiv\d\s\delta(x,y)/\d\s x^a$, 
which transforms as a ``vector density''. It is clear that the Poisson 
brackets of the constraints vanish on the constraint hypersurface because 
the right-hand sides of (\ref{e31})--(\ref{e33}) all vanish there, i.e., 
when $H_a(x)=0=H(x)$ for all $x\in{\cal S}$. This vanishing property of the 
Poisson brackets is characteristic of {\it first-class constraints}. Often, 
the Poisson bracket structure of constraints is that of a Lie bracket in 
which case such constraints are referred to as closed first-class 
constraints. In order for the Poisson bracket structure to be that of a 
Lie bracket, it is necessary that the coefficients of the constraints on 
the right-hand side involve no dynamical variables. Taken by themselves, 
we note from (\ref{e31}) that the three diffeomorphism constraint functions 
form a set of closed first-class constraints. However, because of the last 
equation, (\ref{e33}), it is clear that the complete set of four 
gravitational constraint functions do {\it not} have a Poisson structure 
which is that of a Lie algebra, and consequently the gravitational 
constraints are said to form an open first-class system of constraints. 
Such a situation does not automatically imply trouble in the corresponding 
quantum theory, but significant difficulties do arise in a number of cases. 
Quantum gravity is one of those cases.

Let us proceed formally in order to see the essence of the problem. Suppose 
that $\H_a(x)$ and $\H(x)$ represent local self-adjoint constraint operators 
for the gravitational field. Standard calculations lead to the commutation 
relations (with $\hbar=1$)
 \bn  && [\H_a(x),\H_b(y)]=i\s[\delta_{,a}(x,y)\,\H_b(x)-\delta_{,b}(x,y)\,
\H_a(x)]\;, \\
    && \hskip.18cm[\H_a(x),\H(y)]=i\s\delta_{,a}(x,y)\,\H(x)\;,  \\
 && \hskip.36cm[\H(x),\H(y)]= i\s\half\s\delta_{,a}(x,y)[\,\hg^{ab}(x)\,
\H_b(x)+\H_b(x)\,\hg^{ab}(x)\,]\;,  \en
where to ensure the Hermitian character we have symmetrized the right-hand 
side of the last expression. In the usual Dirac approach to constraints 
alluded to in Sec.~I, one asks that $\Phi_\a\s|\psi\>_{phys}=0$ for all 
constraints. If we assert that $\H_a(x)\s|\psi\>_{phys}=0$ and 
$\H(x)\s|\psi\>_{phys}=0$, then consistency holds for the first two sets 
of constraint commutators, but not for the third commutator in virtue of 
the fact that it is almost surely the case that $\hg^{ab}(x)\,|
\psi\>_{phys}\not\in{\frak H}_{phys}$, even if it were smeared. The 
expected behavior is somewhat like that of the single degree-of-freedom 
example where $Q\s|\psi\>=0$ and $P\s|\psi\>=0$ imply for standard 
Heisenberg operators that $[Q,P]\s|\psi\>=i\s|\psi\>=0$, i.e., there are 
no nonvanishing solutions. This behavior is characteristic of second-class 
constraints, and as a consequence of our discussion we are led to conclude, 
from a quantum mechanical standpoint, that part of the gravitational 
constraints are {\it second-class constraints} \cite{JIMAND}. For the 
projection operator method of constrained system quantization, however, 
second-class constraints cause no special difficulty and, in particular, 
they are treated in just the same way as first-class constraints, as already 
noted in Sec.~I. $(\!\!(${\bf Remark:} Some researchers prefer to modify 
the theory so as to eliminate the second-class nature of the gravitational 
constraints. Instead, we accept the second-class constraints  for what they 
are.$)\!\!)$

Assuming that the constraint fields $\H_a(x)$ and $\H(x)$ are local 
self-adjoint operators, we could---as one of several different 
alternatives---proceed as follows. Initially, besides the real, orthonormal 
set of test functions $\{h_p(x)\}$ introduced in Sec.~II, let us introduce 
an additional set of real test functions $\{f_{p\,A}^a(x)\}$, 
$A\in\{1,2,3\}$, with the following properties:
 \bn &&b(x)\Sigma_{p=1}^\infty\Sigma_{A=1}^3 f_{p\,A}^a(x)\s f_{p\,A}^b(y)
={\tilde g}^{ab}(x)\s\delta(x,y)\;, \\
  &&\hskip.3cm\tint f_{p\,A}^a(x)\s f_{q\,B\;a}(x)\s b(x)\,d^3\!x=
\delta_{pq}\s\delta_{AB}\;,  \en
where $f_{q\,B\;a}(x)\equiv {\tilde g}_{ab}(x)\s f_{q\,B}^b(x)$. Regarding 
the set of functions $\{f_{p\,A}^a(x)\}$,  the index $A$ is a dreibein 
index, while the index $a$ is a three-space vector index. With the help of 
these sets of expansion functions, let us introduce
 \bn  && \H_{(p)\,A}\equiv\tint f_{p\,A}^a(x)\s\H_a(x)\,d^3\!x\;,  \\
   && \hskip.25cm\H_{(p)}\equiv\tint h_p(x)\s\H(x)\,d^3\!x\;,  \en
each for $1\le p<\infty$, and similarly for other vector and scalar 
functions. In this form as well, part of the constraint operators are 
second class. To accomodate all these constraints we introduce a set of 
projection operators defined for all $P\in\{1,2,3,\ldots\}$ and given by
  \bn  &&\E_P\equiv\E(X^2_P\le\delta(\hbar)^2)\;,  \\
  && X^2_P\equiv \Sigma_{p=1}^P\,2^{-p}\,[\,\Sigma_{A=1}^3\,(\H_{(p)\,A})^2+
\H_{(p)}^2\,]\;.  \en
As defined, the projection operators $\E_P$ are regularized and they serve 
to define regularized physical Hilbert spaces ${\frak H}_{phys}\equiv\E_P\s
{\frak H}$. These regularized physical Hilbert spaces may, in turn, be 
characterized by their own reproducing kernels
 \bn  \<\pi'',g''|\s\E_P\s|\pi',g'\>\;,  \en
and the regularized physical Hilbert spaces themselves may, therefore, be 
represented by the associated reproducing kernel Hilbert spaces.

The final step in the present construction procedure would involve suitable 
limits to remove the regularizations. More familiar procedures to enforce 
the constraints are discussed in Sec.~IV.

\subsection{Functional integral representation for the\\relevant projection 
operators}
In an earlier work \cite{UNIV}, we have presented a very general procedure 
to construct the projection operator $\E(\Sigma_\a\Phi^2_\a\le
\delta(\hbar)^2)$ by means of a universal functional integral procedure. 
In particular, it follows that
 \bn \E(\Sigma_\a\Phi^2_\a\le\delta(\hbar)^2)=\int {\bf T}\,
e^{-i\tint\lambda^\a(t)\s\Phi_\a\,dt}\,\D R(\lambda)\;, \en
where ${\bf T}$ denotes the time-ordering operator, 
$\{\lambda^\a(t)\}_{\a=1}^A$, $0\le t< T$, denotes a set of $c$-number 
``Lagrange multiplier'' functions, and $\D R(\lambda)$ denotes a formal 
measure on such functions. A suitable measure $R$ may be determined as 
follows: First, introduce a Gaussian integral over the set 
$\{\lambda^\a(t)\}$ so that 
 \bn  {\cal N}_\gamma\,\int{\bf T}\,e^{-i\tint\lambda^\a(t)\s\Phi_\a\,dt}\,
e^{(i/4\gamma)\tint\Sigma_\a\s\lambda^\a(t)^2\,dt}\,\D\lambda=e^{-i\gamma\s
\Sigma_\a\Phi^2_\a}\;.  \en
Second, and last, integrate over $\gamma$ according to the rule
  \bn  \lim_{\zeta\ra0^+}\,\lim_{L\ra\infty}\,\int_{-L}^L
\frac{\sin\{\gamma[\delta(\hbar)^2+\zeta]\}}{\pi\gamma}\,e^{-i\gamma\s
\Sigma_\a\Phi^2_\a}\,d\gamma=\E(\Sigma_\a\Phi^2_\a\le\delta(\hbar)^2)\;.  \en 
The inclusion of the variable $\zeta$ and the limit $\zeta\ra0^+$ ensures 
that we include the equality sign in the argument of $\E$. Observe that this 
construction is entirely independent of the nature of the set of constraints 
$\{\Phi_\a\}$. 

Next, we continue to proceed formally in order to envisage how the projection 
operator method may be used
in the case of gravity. Adopting the foregoing analysis, we suggest that 
  \bn &&\E_P=\E(\Sigma_{p=1}^P\,2^{-p}\,[\,\Sigma_{A=1}^3\,(\H_{(p)\,A})^2
+\H_{(p)}^2\,])\no\\  
  &&\hskip.7cm=\int{\bf T}\,e^{\s-i\s\Sigma_{p=1}^P\tint[\s\Sigma_{A=1}^3\s 
N_{(p)\,A}\,\H_{(p)\,A}+N_{(p)}\s\H_{(p)}\s]\,dt}\,\D R(N^a,N)  \en
for an appropriately defined formal measure $R$. Observe that the integral 
over the measure $R$ may well integrate over degrees of freedom that are 
not present in the time-ordered product (such as for $p>P$); however, there 
is no harm in doing so since $R$ is already defined so that $\tint\D R=1$. 
We may also go one step further and assert that as $P\ra\infty$ we obtain 
the formal expression
  \bn \E=\int{\bf T}\,e^{\s-i\s\tint[N^a\s\H_a+N\s\H]\,d^3\!x\,dt}\,
\D R(N^a,N)\;,  \en
which, heuristically at least, realizes the projection operator that 
projects the original Hilbert space onto a correspondingly regularized 
physical Hilbert space. 

\subsection{Functional integral representation of the \\reproducing kernel 
for the physical Hilbert space}
In Sec.~II, we presented in (\ref{i22}) a continuous-time regularized 
functional integral 
representation of the
reproducing kernel $\<\pi'',g''|\pi',g'\>$ for the original Hilbert space. 
The reproducing kernel for the (regularized) physical Hilbert space is 
given, in turn, by the expression $\<\pi'',g''|\s\E\s|\pi',g'\>$. In order
to give this latter expression a functional integral representation we 
first regard
   \bn  \tint[N^a\s\H_a+N\s\H]\,d^3\!x  \en
as a time-dependent ``Hamiltonian'' for some fictitious theory, in which 
case 
 \bn  && \<\pi'',g''|\s{\bf T}\,e^{\s-i\s\tint[N^a\s\H_a+N\s\H]\,d^3\!x\,dt}
\s|\pi',g'\> \no\\
  &&\hskip1cm=\lim_{\nu\ra\infty}{\o{\cal N}}_\nu\,\int\exp\{-i\tint[g_{ab}
\s{\dot\pi}^{ab}+N^a\s H_a+N\s H]\,d^3\!x\,dt\} \no\\
  &&\hskip1.5cm\times\exp\{-(1/2\nu)\tint[b(x)^{-1}g_{ab}g_{cd}{\dot\pi}^{bc}
{\dot\pi}^{da}+b(x)g^{ab}g^{cd}{\dot g}_{bc}{\dot g}_{da}]\,d^3\!x\,dt\}\no\\
  &&\hskip2cm\times\Pi_{x,t}\,\Pi_{a\le b}\,d\pi^{ab}(x,t)\,dg_{ab}(x,t)\;. 
\label{f35} \en
In this expression, there appear symbols $H_a(\pi,g)$ and $H(\pi,g)$ 
corresponding to the quantum operators $\H_a$ and $\H$. Superficially, 
these symbols may (formally) be identified with the classical 
diffeomorphism and Hamiltonian constraint functions, in which case the 
expression (\ref{f35}) contains in its phase, and up to a surface term, 
the full Einstein action. Thus (\ref{f35}) comes ever closer to looking 
like a more traditional functional integral for gravity.

However, before integrating over the functions $N^a$ and $N$ and completing 
the story, we need to caution the reader that $\hbar$ is not zero (but 
rather one) and therefore the symbols $H_a$ and $H$ may not coincide with 
their usual classical expressions. All we can say at present is that $H_a$ 
is a symbol for the operator $\H_a$, $a=1,2,3$, and that $H$ is a symbol for 
the operator $\H$. The connection between symbol and operator is implicitly 
contained in (\ref{f35}), and since, for the moment, the functions $N_a$ and 
$N$ are general functions within our control, we may use that fact to assert 
that
 \bn &&\hskip-.4cm\<\pi'',g''|\s\tint[M^a(y)\s\H_a(y)+M(y)\s\H(y)]\,d^3\!y
\s|\pi',g'\> \no\\
  &&\hskip.6cm=\lim_{\nu\ra\infty}{\o{\cal N}}_\nu\,\int e^{-i\tint g_{ab}
\s{\dot\pi}^{ab}\,d^3\!x\,dt}
\,\tint[\s M^a(y)\s H_a(y,s)+M(y)\s H\s(y,s)]\,d^3\!y \no\\
  &&\hskip1.2cm\times\exp\{-(1/2\nu)\tint[b(x)^{-1}g_{ab}g_{cd}{\dot\pi}^{bc}
{\dot\pi}^{da}+b(x)g^{ab}g^{cd}{\dot g}_{bc}{\dot g}_{da}]\,d^3\!x\,dt\}\no\\
  &&\hskip1.7cm\times\Pi_{x,t}\,\Pi_{a\le b}\,d\pi^{ab}(x,t)\,dg_{ab}(x,t)  
\label{f36}\en
for any smooth (test) functions $M^a$ and $M$ and for any time $s$, $0<s<T$. 
Equation (\ref{f36}) gives an implicit connection between symbol and 
operator for the present theory. We observe that the more traditional 
connection between symbol and operator that normally holds for 
Wiener-regularized coherent-state path integrals \cite{DAUB} 
is unavailable in the present case since we are dealing with so-called 
weak coherent states for which no resolution 
of unity exists; see \cite{P-I,GUTZ}. 

In addition, thanks to analyticity in the present case, the diagonal matrix 
elements of an operator uniquely determine the operator, and so we can also 
assert the connection between symbol and operator (given by setting 
$\pi'',g''=\pi',g'$),
\bn &&\hskip-.4cm\tint[M^a(y)\<\pi',g'|\H_a(y)|\pi',g'\>+M(y)
\<\pi',g'|\H(y)|\pi',g'\>]\,d^3\!y  \no\\
  &&\hskip.6cm=\lim_{\nu\ra\infty}{\o{\cal N}}_\nu\,\int 
e^{-i\oint g_{ab}\s{\dot\pi}^{ab}\,d^3\!x\,dt}\,\tint[\s M^a(y)\s 
H_a(y,s)+M(y)\s H\s(y,s)]\,d^3\!y \no\\
  &&\hskip1.2cm\times\exp\{-(1/2\nu)\tint[b(x)^{-1}g_{ab}g_{cd}
{\dot\pi}^{bc}{\dot\pi}^{da}+b(x)g^{ab}g^{cd}{\dot g}_{bc}{\dot g}_{da}]\,
d^3\!x\,dt\}\no\\
  &&\hskip2cm\times\Pi_{x,t}\,\Pi_{a\le b}\,d\pi^{ab}(x,t)\,dg_{ab}(x,t)\;,  
\label{f37}\en
again for smooth functions $M^a$ and $M$ and any $s$, $0<s<T$. We note that 
$\<\pi',g'|\H_a(y)|\pi',g'\>$ and $\<\pi',g'|\H(y)|\pi',g'\>$ denote still 
other symbols that are often associated with the local operators $\H_a(x)$ 
and $\H(x)$, respectively. In (\ref{f37}), the notation $\oint g_{ab}\s{\dot
\pi}^{ab}\,d^3\!x\,dt$ means that only closed paths in phase space enter, 
i.e., just those paths for which the functions 
   \bn && \pi^{ab}(x,0)=\pi^{ab}(x,T)\equiv\pi'^{ab}(x) \;,\\
       &&g_{ab}(x,0)=g_{ab}(x,T)\equiv g'_{ab}(x)   \label{f38}\en
for all $x\in{\cal S}$. Note that a {\it closed} line integral in phase 
space involves just the symplectic form, and the result of the integral 
$\oint g_{ab}\s{\dot\pi}^{ab}\,d^3\!x\,dt$ is {\it invariant} under any 
(smooth) change of canonical coordinates. 

\subsubsection*{Reproducing kernel for the physical Hilbert space}
We now complete the story by interpreting the otherwise arbitrary $c$-number 
functions $N^a$ and $N$ as Lagrange multiplier functions and integrating 
them out of (\ref{f35}). Since $N^a$ and $N$ are not dynamical variables 
that must enter the formal phase-space functional integral measure in a 
prescribed way (i.e., as ``$dp\, dq$''), we are free to integrate them as we 
choose---and we choose to integrate them in such a way as to {\it enforce 
the quantum constraints}, at least in a regulated fashion. As explained 
above, one natural way to achieve our goal involves the formal integration 
measure $\D R(N^a,N)$. 

Combining several steps previously described, we now assert that the 
reproducing kernel for the regularized physical Hilbert space has the 
phase-space functional integral representation given by
  \bn  && \<\pi'',g''|\s\E\s|\pi',g'\>  \no\\
  &&\hskip1cm=\int \<\pi'',g''|{\bf T}\,e^{-i\tint[\s N^a\s\H_a+N\s\H\s]
\,d^3\!x\,dt}\s|\pi',g'\>\,\D R(N^a,N)\no\\
 &&\hskip1cm=\lim_{\nu\ra\infty}{\o{\cal N}}_\nu\s\int 
e^{-i\tint[g_{ab}{\dot\pi}^{ab}+N^aH_a+NH]\,d^3\!x\,dt}\no\\
  &&\hskip1.5cm\times\exp\{-(1/2\nu)\tint[b(x)^{-1}g_{ab}g_{cd}
{\dot\pi}^{bc}{\dot\pi}^{da}+b(x)g^{ab}g^{cd}{\dot g}_{bc}{\dot g}_{da}]\,
d^3\!x\,dt\}\no\\
  &&\hskip2cm\times\bigg[\Pi_{x,t}\,\Pi_{a\le b}\,d\pi^{ab}(x,t)\,
dg_{ab}(x,t)\bigg]\,\D R(N^a,N)\;. \label{f39}\en
In this final expression we have reached our primary goal, at least from a 
formal perspective. Despite the general appearance of ({\ref{f39}), we 
emphasize once again that this representation has been based on the affine 
commutation relations and {\it not} on any canonical commutation relations! 
Later, we shall discuss a more careful definition of this formal expression 
along lines introduced in Sec.~II, but for now let us examine (\ref{f39}) 
for its own sake.  

We first comment on the range of integration for the ``lapse'' variable 
$N(x,t)$, a subject of recurrent interest \cite{TEIT}. Our view is that 
the range $-\infty< N(x,t)<\infty$ is the proper range when quantizing the 
theory. After all, in the Hamiltonian viewpoint, {\it space-time is a 
derived structure} of the classical theory. In principle, the issue here 
is no more complicated than for the reparametrized one-dimensional free 
particle. For this example, the original classical action is 
$I=\tint[p\s{\dot q}-\half p^2]\s dt$, where ${\dot q}\equiv dq/dt$, 
and has solutions $p(t)=p_o$ and $q(t)=p_ot+q_o$. The reparametrized 
version is given by
$I'=\tint[p\s q^*+s\s t^*-\lambda(s+\half p^2)]\s d\tau$, where 
$q^*\equiv dq/d\tau$, $t^*\equiv dt/d\tau$, and has solutions $p(\tau)=p_o$, 
$s(\tau)=s_o=-\half p^2_o$, $t(\tau)\equiv\int_0^\tau\,\lambda(\sigma)\s 
d\sigma$, and $q(\tau)=t(\tau)p_o+q_o$. The function $\lambda(\tau)$ is 
essentially arbitrary. If, for example, $\lambda(\tau)=3\tau^2-1$, i.e., 
$t(\tau)=\tau(\tau^2-1)$, the solution seems ``to go backward in time'', 
but that interpretation gives to the variable $\tau$ an unwarranted physical 
significance. The given solution is not wrong, it just repeats itself for a 
while. We can avoid a repeating behavior, e.g., by simply dropping the 
interval $-1\le\tau<1$. No such issues occur if we require that 
$\lambda(\tau)>0$ for all $\tau$, in which case $\tau$ does indeed merit 
the name of ``reparametrized time''. By analogy, the function 
$N(x,t)$---which we have loosely called the lapse function---only deserves 
that name when, in the classical solution space, we insist that $N(x,t)>0$; 
otherwise it is just another Lagrange multiplier function, no more and 
no less. 

When one starts from a classical perspective, with its focus on physically 
relevant functions $N(x,t)$ which are strictly positive, it is a conceptual 
leap to change to functions $N(x,t)$ that can take on both signs \cite{TEIT}. 
However, when one starts from the quantum theory, as we have done, there is 
no such leap to make.

As a second topic regarding (\ref{f39}) we focus on its general structure. 
With $H_a$ and $H$ formally equal to the constraint functions of gravity 
(possibly up to terms in $\hbar$), the action appearing in the phase factor 
is indeed appropriate to gravity \cite{TEIT}. Moreover, the domain of 
integration is restricted to positive-definite metrics $\{g_{ab}(x,t)\}>0$. 
Indeed, the particular $\nu$-dependent regularizing phase-space metric in 
(\ref{f39}) {\it prevents the metric variable $\{g_{ab}\}$ from escaping the 
positive-definite domain}.

As a useful analogy, we note that the two-dimensional phase space metric 
 \bn  \beta^{-1} q^2\s dp^2+\beta\s q^{-2}\s dq^2  \en
is geodesically complete in the half-space $(p,q)\in {\mathbb R}
\times{\mathbb R}^+$, and when it is part of a Brownian motion measure, as 
in the formal expression [cf., (\ref{g25})]
  \bn {\cal N}\, e^{-(1/2\nu)\tint[\beta^{-1}\s q^2\s{\dot p}^2+\beta\s 
q^{-2}\s{\dot q}^2]\,dt}\,\D p\,\D q\;, \en
it automatically restricts the Brownian motion trajectories to the 
half-space ${\mathbb R}\times{\mathbb R}^+$.

Lastly, we comment on the formal functional integration measure in 
(\ref{f39}), specifically for the Lagrange multiplier functions $N^a$ and 
$N$. The formal measure $\D R$ has been defined earlier and is unlike 
conventional measures chosen for such variables. As emphasized here, and 
elsewhere \cite{UNIV}, the measure $\D R$ is {\it designed to implement the 
quantum constraints}---as befits a quantum theory---and it has {\it not} 
been selected to enforce the classical constraints. Observe well that we 
have not ``blindly postulated'' the functional integral (\ref{f39}), but 
instead it has been {\it derived} as a specific functional integral 
representation of the well-chosen quantum matrix elements on the left-hand 
side. Unfamiliar as the measure $\D R$ may appear, we maintain that 
$\D R$---or some other measure equivalent to it---is the proper measure to 
choose to achieve our goal. Whether different treatments of the Lagrange 
multiplier functions that have been adopted by other workers are indeed 
equivalent or not to the use of $\D R$ is an interesting question, but it 
is not one we pursue here.

\section{\sc DISCUSSION}
In the preceding analysis, we have been strongly guided by the operator 
structure of the assumed theory of affine quantum gravity, and this 
discussion has led us to the formal functional integral representation 
({\ref{f39}) for the desired matrix elements. As was previously the case 
[cf., (\ref{e20})], we now wish to turn (\ref{f39}) around and adopt the 
formal functional integral as our starting point and, in effect, use that 
expression to {\it define} the reproducing kernel for the regularized 
physical Hilbert space. Specifically, for that purpose, let us focus on the 
formal expression
\bn 
 &&\lim_{\nu\ra\infty}{\o{\cal N}}_\nu\s\int e^{i\tint[{\pi}^{ab}
{\dot g}_{ab}-N^aH_a-NH]\,d^3\!x\,dt}\no\\
  &&\hskip1.5cm\times\exp\{-(1/2\nu)\tint[b(x)^{-1}g_{ab}g_{cd}
{\dot\pi}^{bc}{\dot\pi}^{da}+b(x)g^{ab}g^{cd}{\dot g}_{bc}{\dot g}_{da}]\,
d^3\!x\,dt\}\no\\
  &&\hskip2cm\times \D\pi\,\D g\,\D R(N^a,N)\;. \label{g45}\en
Note the change of the kinematic term, which simply amounts to a phase 
factor in the definition of the coherent states. Also we have introduced 
the common shorthand $\D \pi\,\D g$ for the bracketed term in (\ref{f39}).
For the sake of discussion, we shall refer to (\ref{g45}) as the 
``nonstandard expression''. Our goal in this section is to discuss the 
nonstandard expression and see what steps are necessary to give it a proper 
meaning. We shall do so in a three step procedure: First, we compare the 
``standard'' (see below) and ``nonstandard'' expressions. Second, we discuss 
a regularization and its removal that ultimately involves the elimination 
of the scalar density $b(x)$. Third, we examine the aspect of the problem 
that normally accounts for the perturbative nonrenormalizability of the 
gravitational field. 

\subsection{First look at the Nonstandard Expression}
It is interesting to compare (\ref{g45}) with what we refer to as the 
standard expression for a phase-space functional integral for gravity. 
By the ``standard expression'' we mean the formal functional integral
\bn  {\cal M}\int e^{\s i\s\tint[\pi^{ab}{\dot g}_{ab}-N^a\s H_a-N\s H]\,
d^3\!x\,dt}\,\D\pi\,\D g\,\D N  \label{g46}\;, \en
where $\D N\equiv\Pi_{x,t}\s dN(x,t)\,\Pi_a\s dN^a(x,t)$.
In several important ways, the standard expression is {\it very} different 
than the nonstandard expression. Let us first comment on some of those 
differences. Much as it would be nice to think otherwise, one must recognize 
that the standard expression (\ref{g46}) is {\it totally undefined} as it 
stands; it is little more than a fancy way of writing $0\times(\infty)$. 
It begs for a definition as the limit of meaningful expressions 
[much as $0\times(\infty)$ may, for example, be defined as $\lim_{x\ra0}\s 
x\times(7/x)=7$], but what set of meaningful expressions should be chosen 
in the gravitational case is far from clear. A lattice limit? But then, 
what form should the regularized lattice expressions take? Symmetry and 
covariance offer only limited guidance. In point of fact, this same question 
faces any standard phase-space path integral, even that for a single degree 
of freedom in which a conventional lattice definition---as originally 
envisaged by Tobocman \cite{TOBOC}---makes certain assumptions about the 
nature of the phase-space coordinates which may or may not be true. 
$[\!\![${\bf Remark:} The skeptical reader is urged to propose a lattice 
prescription to quantize the nonrelativistic free particle of unit mass by 
a phase-space path integral whose Hamiltonian $H$ is expressed in canonical 
coordinates $p$ and $q$ such that 
$H=\half(p^2+q^2)$.$]\!\!]$

A further complication of the standard functional integral expression 
(\ref{g46}) arises from the {\it unbounded nature} of the formal integral 
$\tint\cdots\D N$ over the Lagrange multiplier variables. This choice of 
integration measure, which is designed to enforce the classical constraints 
and thereby reduce the classical phase space before quantization, 
necessitates {\it gauge fixing} (to eliminate concomitant divergences) 
which reduces the classical phase space further to the physical phase space 
(at least locally) where each point labels a physically distinct state. 
Quantization on the reduced phase space is formally aided by the introduction 
of some additional factor (e.g., a Faddeev-Popov determinant, or its 
analogue), which may well lead to significant (Gribov) ambiguities which 
require a substantial modification of the functional integral 
\cite{TMF,SHAB}, and give rise to serious problems (such as unitary 
violation) within a BRST formulation \cite{SF}.

\subsubsection*{Nonstandard expression}
Let us raise similar issues regarding the nonstandard expression (\ref{g45}). 
Although (\ref{g45}) is formal as it stands, it can, to a considerable 
extent, already be regarded as ``nearly'' well defined. As noted in Sec.~II, 
we can combine several factors together to make, for all finite $\nu$, a 
positive, countably-additive, pinned measure on generalized functions. 
What makes (\ref{g45}) not well defined is the fact that the formal 
integrand does not constitute an integrable function with respect to that 
measure. We have already encountered that problem in Sec.~II before any 
constraints were introduced, and we found that we could overcome that problem 
by regularizing the integrand and removing that regularization as a final 
step. Superficially, the same property holds when the constraint functions 
are present (say for fixed Lagrange multiplier values), save for one very 
important distinction (involving the field operator representation) which we 
shall address in the second point of discussion below.

Regarding the integration over the Lagrange multiplier variables, we 
emphasize the vast difference afforded by the projection operator method. 
First and foremost is the fact that {\it no gauge fixing} is introduced, no 
ghosts are used, no
Faddeev-Popov determinant (or its analogue) arises, and consequently, no 
Gribov ambiguities can exist. These properties arise, largely, because 
$\tint\D R(N^a,N)=1$, while $\tint\D N=\infty$. The difference here could 
not be greater, and it arises because in the former case one quantizes 
first and reduces second, while in the latter case one reduces first and 
quantizes second. Except in basically trivial cases, the second option is 
fraught with substantial obstacles \cite{SHAB}.

One of the most significant differences between the standard and the 
nonstandard expressions refers to the {representation of the quantum 
mechanical amplitudes} that is involved. For the standard expression, it is 
usually assumed that (\ref{g46}) leads to a representation in which the 
metric field operator $\hg_{ab}(x)$ is diagonalized and thus is sharply 
represented. In combination with any needed auxiliary factor in the 
functional integral, diffeomorphism invariance suggests that (\ref{g46}) 
depends only on the ``geometry'' of the initial and final three-surfaces, 
and not on the details of any specific metric expressions \cite{MTW,TEIT}. 
An analogous view is also prominent in the associated 
``loop quantum gravity'' in which, e.g., bras and kets depend only on knot 
invariants as labels of ``physically'' distinguishable states \cite{ASHT}. 
In contrast, the representation afforded by the nonstandard expression 
(\ref{g45}) is that of a {\it coherent-state representation}, which depends 
on smooth metric $g_{ab}$ and momentum $\pi^{ab}$ fields, that represent 
not {\it sharp} operator (eigen)values but {\it mean} operator values, e.g., 
  \bn  && \<\pi,g|\s\hp^{ab}(x)\s|\pi,g\>=\pi^{ab}(x)\;,  \\   
   && \hskip.07cm\<\pi,g|\s \hg_{ab}(x)\s|\pi,g\>=g_{ab}(x)\;,  \\ 
   &&\hskip.14cm\<\pi,g|\s{\hat\pi}^b_a(x)\s|\pi,g\>=g_{ac}(x)\s
\pi^{cb}(x)\;. \en
Note well, that besides the local self-adjoint metric $\hg_{ab}(x)$ and 
scale field $\hp^b_a(x)$, we have used the local {\it symmetric} 
(but {\it not} self-adjoint) momentum operator $\hp^{ab}(x)$ in the first 
of these expressions. These expectations are not gauge invariant, nor 
should they be, since they are taken in the ``original'' Hilbert space 
where the constraints are not fulfilled. The gauge invariant part of the 
metric field, for example, and in so far as the regularized physical Hilbert 
space is concerned, is determined by the matrix elements
  \bn  \<\pi'',g''|\s\E\s g_{ab}(x)\s\E\s|\pi',g'\>  \;,  \en
which is an expression that does not require restricting the functional 
dependence of the bras and kets.  

\subsection{Second Look at the Nonstandard Expression}
In interpreting (\ref{g45}) we have concluded above that we must first 
regularize the integrand in order to obtain an integrable function. For the 
kinematic term $\tint {\pi}^{ab}\s{\dot g}_{ab}\,d^3\!x\,dt$---and even for 
the diffeomorphism constraint contribution $-\tint N^aH_a\,d^3\!x\,dt$---any 
natural regularization, such as one based on the expansion functions 
$\{h_p(x)\}$ and $\{f_{p\,A}^a(x)\}$, or a lattice formulation as discussed 
in Sec.~II, will be compatible regularizations. For these terms alone, the 
limit of the regularized functional integral as the regularization is 
removed will converge to the desired result. The implication of this fact 
is that these parts of the integrand are compatible with the initial 
(ultralocal) representation of the field operators; in fact, they are 
compatible with any diffeomorphism invariant realization. However, when 
it comes time to consider the Hamiltonian constraint, the behavior is 
quite different. While it is true that regularizing the Hamiltonian 
constraint will lead to a set of well-defined functional integrals, the 
limit of such regularized expressions will {\it not} converge to an 
acceptable result. There are two basic and important reasons for this 
unsatisfactory behavior, one of which (wrong field operator representation) 
we will deal with in this section, the other of which (perturbative 
nonrenormalizability of gravity) we will discuss in the next section.

The first reason for the lack of a suitable convergence of the regularized 
nonstandard functional integral relates to the fact that the representation 
of the field operators needed to satisfy the Hamiltonian constraint is 
unitarily {\it in\/}equivalent to the ultralocal representation imposed 
in the initial stage of the quantization procedure. It is at the present 
stage of the analysis that we finally encounter the fact that our initial 
choice of field operator representation is incompatible with making the 
Hamiltonian constraint operator $\H(x)$ into a densely defined local 
operator. Stated otherwise, using the ultralocal operator representation, 
the operator $\tint N(x,t)\s\H(x)\,d^3\!x\,dt$, for any nonzero smooth 
function $N(x,t)$, has only the zero vector in its domain. This defect must 
be fixed before proceeding, and in so doing we will be explicitly led to a 
new representation of the field operators, one that is unitarily inequivalent 
to our starting (ultralocal) representation. In the process of effecting 
this change of representation, the scalar density $b(x)$ will disappear from 
the scene entirely!

\subsubsection*{Pedagogical example}
It is pedagogically useful to outline an analogous story for a simpler and 
more familiar example. Consider general, locally self-adjoint field and 
momentum operators, $\hph(x)$ and $\hp(x)$, $x\in{\mathbb R}^3$, which 
satisfy the canonical commutation relations
  \bn  [\hph(x),\hp(y)]=i\s\delta(x-y)\;.  \label{h73}\en
Build a set of coherent states
 \bn  |\pi,\phi\>\equiv e^{i\s[\hph(\pi)-\hp(\phi)]}\,|\eta\>\;,  \en
where $\hph(\pi)\equiv\tint\hph(x)\s\pi(x)\,d^3\!x$ and $\hp(\phi)\equiv
\tint\hp(x)\s\phi(x)\,d^3\!x$, with
$\pi$ and $\phi$ real test functions, and $|\eta\>$ is a normalized but 
otherwise unspecified fiducial vector. Note well  that the choice of 
$|\eta\>$ in effect determines the representation of the canonical field 
operators. We next present a portion of the story from Ref.~8.

We initially choose $|\eta\>$ to correspond to an ultralocal representation 
such that
  \bn  && \<\pi'',\phi''|\pi',\phi'\>=\exp\{\half\s i\s\tint[\phi''(x)\s
\pi'(x)-\pi''(x)\s\phi'(x)]\,d^3\!x\}\no\\
  &&\hskip-.8cm\times\exp(\!\!(-\quarter\tint\{M(x)^{-1}
[\pi''(x)-\pi'(x)]^2+M(x)[\phi''(x)-\phi'(x)]^2\}\,d^3\!x\,)\!\!)\;.  \en
Here $M(x)$, $0<M(x)<\infty$, is an arbitrary (smooth) function of the 
ultralocal representation with the dimensions of {\sf M}. The given 
ultralocal field operator representation is in fact unitarily inequivalent 
for each distinct function $M(x)$. [Note well that $M(x)$ here plays the 
role of $b(x)$ in the present paper.]

We wish to apply this formulation to describe the {\it relativistic free 
field of mass $m$} for which the Hamiltonian operator is formally given by
  \bn  \H=\half\s\tint:\{\hp(x)^2+[\nabla\hph(x)]^2+m^2\s\hph(x)^2\}:\,
d^3\!x\;.  \en
If we build this operator out of the field and momentum operators in the 
ultralocal representation, then no matter what vector is used to define 
$:\;:$, $\H$ will have only the zero vector in its domain. We need to change 
the field operator representation, which means we have to change the 
fiducial vector from $|\eta\>$ to $|0;m\>$, the true ground state of the 
proposed Hamiltonian operator $\H$. 

Let us first regularize the formal Hamiltonian $\H$. To that end, let 
$\{u_n(x)\}$ denote a complete set of real, orthonormal functions on 
${\mathbb R}^3$ and define the sequence of kernels, for all $N\in\{1,2,3,
\ldots\}$, given by
  \bn  K_N(x,y)\equiv\sum_{n=1}^N\,u_n(x)\s u_n(y)\;,  \en
which converges to $\delta(x-y)$ as a distribution when $N\ra\infty$. Let
  \bn  &&  \hph_N(x)\equiv \tint K_N(x,y)\s\hph(y)\,d^3\!y\;, \\
  &&   \hp_N(x)\equiv\tint K_N(x,y)\s\hp(y)\,d^3\!y\;,  \en
and with these operators build the sequence of regularized Hamiltonian 
operators
  \bn  \H_N\equiv\half\s\tint :\{\hp_N(x)^2+[\nabla\hph_N(x)]^2+
m^2\hph_N(x)^2\}:\,d^3\!x  \en
for all $N$, where $:\;:$ denotes normal order with respect to the ground 
state $|0;m\>_N$ of $\H_N$.

We would like to have a constructive way to identify the ground state of 
$\H_N$. For this purpose consider the set
\bn  S_N\equiv\bigg\{\frac{\Sigma_{j,k=1}^J\,a^*_j\s a_k\s\<\pi_j,\phi_j|\s 
e^{-\H_N^2}\s|\pi_k,\phi_k\>}{\Sigma_{j,k=1}^J\,a^*_j\s a_k\s\<\pi_j,\phi_j|
\pi_k,\phi_k\>}\,:\,J<\infty\bigg\}  \en
for general sets $\{a_j\}$ (not all zero), $\{\pi_j\}$, and $\{\phi_j\}$. 
(How these expressions may be generated is discussed in Ref.~8.) As $N$ 
grows, the general element in $S_N$ becomes exponentially small, save for 
elements that correspond to vectors which well approximate the ground state 
$|0;m\>_N$. Suitable linear combinations can convert the original reproducing 
kernel $\<\pi'',\phi''|\pi',\phi'\>$ to the reproducing kernel 
 $\<\pi'',\phi'';m|\pi',\phi';m\>_N$ which is based on a fiducial vector 
that has the form $|0;m\>_N$ for the first $N$ degrees of freedom and is 
unchanged for the remaining degrees of freedom. Finally, we may take the 
limit $N\ra\infty$ which then leads to
 \bn && \<\pi'',\phi'';m|\pi',\phi';m\>=\exp\{\half\s i\s
\tint[{\tilde\phi}''^*(k)\s{\tilde\pi}'(k)-{\tilde\pi}''^*(k)\s
{\tilde\phi}'(k)]\,d^3\!k\}\no\\
  &&\hskip-.6cm\times\exp(\!\!(-\quarter\tint\{\omega(k)^{-1}|
{\tilde\pi}''(k)-{\tilde\pi}'(k)|^2+\omega(k)|{\tilde\phi}''(k)-{\tilde
\phi}'(k)|^2\}\,d^3\!k\s)\!\!)\;,  \en
where $\omega(k)\equiv\sqrt{k^2+m^2}$ and ${\tilde\pi}(k)
\equiv(2\pi)^{-3/2}\tint e^{-ik\cdot x}\,\pi(x)\,d^3\!x$, etc. The 
procedure sketched above is referred to as {\it recentering the coherent 
states} or equivalently as {\it recentering the reproducing kernel}. This 
form of reproducing kernel is no longer ultralocal and contains no trace of 
the scalar function $M(x)$, whatever form it may have had. Moreover, and 
this is an important point, the new representation is fully compatible with 
the Hamiltonian $\H$ being a nonnegative, self-adjoint operator. Indeed, 
the expression for the propagator is given by
\bn  \<\pi'',\phi'';m|\s e^{-i\H\s T}\s|\pi',\phi';m\> 
  =L''\s L'\s\exp[\tint{\tilde\zeta}''^*(k)\s e^{-i\s\omega(k)\s T}
\s{\tilde\zeta}'(k)\,d^3\!k\s]\;, \label{h83}\en
where
  \bn  && {\tilde\zeta}(k)\equiv[\s\omega(k)^{1/2}\s{\tilde\phi}(k)+i\s
\omega(k)^{-1/2}\s{\tilde\pi}(k)]/\sqrt{2}\;,  \\
   && \hskip.7cm L\equiv \exp[-\half\tint|{\tilde\zeta}(k)|^2\,d^3\!k\s]\;. 
\en
The definition offered by (\ref{h83}) is continuous in $T$, which is the 
principal guarantor that the expression
  \bn \H=\half\tint:\{\hp(x)^2+[\nabla\hph(x)]^2+m^2\s\hph(x)^2\}:
\,d^3\!x\;,  \en
where $:\;:$ denotes normal ordering with respect to the ground state 
$|0;m\>$ of the operator $\H$,
 is a self-adjoint operator as desired.

Let us summarize the basic content of the present pedagogical example. 
Even though we started with a very general ultralocal representation, as 
characterized by the general function $M(x)$, we have forced a complete 
change of representation to one compatible with the Hamiltonian operator 
for a relativistic free field of arbitrary mass $m$. In so doing all trace 
of the initial arbitrary function $M(x)$ has disappeared, and in its place, 
effectively speaking, has appeared the pseudo-differential operator 
$\sqrt{-\nabla^2+m^2}$ having only its dimension (mass) in common with the 
original function $M(x)$. The original ultralocal representation is 
completely gone!
$(\!\!(${\bf Remark:} A moments reflection should convince the reader that 
a comparable analysis can be made for either the interacting $\phi^4_2$ or 
$\phi^4_3$ model as well, both of which satisfy (\ref{h73}), showing that 
the general argument is not limited just to free theories; see 
\cite{REL}.$)\!\!)$ 

\subsubsection*{Strong coupling gravity}
The discussion in the present paper has been predicated on the assumption 
that we are analyzing the gravitational field and therefore the classical 
Hamiltonian is that given in (\ref{h45}). However, it is pedagogically 
instructive if we briefly comment on an approximate theory---based on the 
so-called ``strong coupling approximation'' \cite{PIL}---where the 
Hamiltonian constraint (\ref{h45}) is replaced by the expression
  \bn  H_{SCA}(x)\equiv g(x)^{-1/2}[\,\pi^a_b(x)\s\pi^b_a(x)-
\half\pi^a_a(x)\s\pi^b_b(x)\,]+2\Lambda\s g(x)^{1/2}\;, \label{j91}\en
in which the term proportional to $^{(3)}R(x)$ has been dropped, and 
where we have also introduced the cosmological constant $\Lambda$ (with 
dimensions ${\sf L}^{-2}$) and an associated auxiliary term in the 
Hamiltonian. The result is a model for which the new Hamiltonian 
constraint (\ref{j91}) is indeed compatible with some form of an 
ultralocal representation. The proper form of that ultralocal 
representation may be determined by a similar procedure, i.e., by studying 
an analogue of the set $S_N$, and by recentering the reproducing kernel 
based on ensuring that the quantum version of $H_{SCA}(x)$ is a local 
self-adjoint operator. In so doing, we note that it may happen that not 
all arbitrariness of the original scalar density $b(x)$ is 
``squeezed out'' by the recentering procedure described above. This 
situation occurs because a one-parameter arbitrariness generally remains 
for typical ultralocal models \cite{KBOOK}. (Any remaining arbitrariness 
in the ultralocal case is in contrast with that of the true Hamiltonian 
constraint for gravity for which we expect no trace of the original 
function $b(x)$ to remain.) These remarks conclude our discussion of 
strong coupling gravity.

\subsection{Third Look at the Nonstandard Expression}
The preceding discussion has been based on the assumption that some 
fiducial vector can be found compatible with the Hamiltonian operator 
constraint, or stated otherwise, that the Hamiltonian constraint can 
actually be realized as a local self-adjoint operator. This requirement 
is by no means obvious, and it is to this issue that we now turn our 
attention. The difficulty arises because the naive form of the Hamiltonian 
constraint operator almost surely needs some form of renormalization if 
it is going to be well defined. If perturbation theory is any guide, we 
not only expect that there will be renormalization counterterms, but 
because gravity is perturbatively nonrenormalizable, one may expect an 
infinite number of distinct counterterms. On the other hand, as we next 
argue, it is possible that perturbation theory is not a very reliable 
guide in the case of perturbatively nonrenormalizable theories.

\subsubsection*{Nonrenormalizable scalar fields}}
Consider the case of perturbatively nonrenormalizable quartic, 
self-interact-ing scalar fields, i.e., the so-called $\phi^4_n$ theories, 
where the space-time dimension $n\ge5$. On the one hand, viewed 
perturbatively, such theories entail an infinite number of distinct 
counterterms. On the other hand, the continuum limit of a straightforward 
Euclidean lattice formulation leads to a quasifree theory---a genuinely 
{\it noninteracting} theory---whatever choice is made for the renormalized 
field strength, mass, and coupling constant \cite{SOK}. In the author's view 
both of these results are unsatisfactory. Instead, it is possible that an 
{\it intermediate behavior} holds true, even though that cannot be 
proven yet. Let us illustrate an analogous but simpler situation where 
the conjectured intermediate behavior can be rigorously established.

Consider an ultralocal quartic interacting scalar field, which, viewed 
classically, is nothing but the relativistic $\phi^4_n$ model with all 
the spatial gradients in the usual free term dropped. As a mathematical 
model of quantum field theory, an ultralocal model is readily seen to be 
perturbatively nonrenormalizable, while the continuum limit of a 
straightforward lattice formulation becomes quasifree, basically because 
of the vise grip of the Central Limit Theorem. Perturbative 
nonrenormalizability and lattice-limit triviality is similar to the 
behavior for relativistic $\phi^4_n$ models, but for the simpler 
ultralocal model for which an intermediate approach can be rigorously 
proven to hold \cite{KBOOK}. Roughly speaking, a characterization of this 
intermediate behavior is the following: From a functional integral 
standpoint, and for any positive value of the quartic coupling constant, 
the quartic interaction acts like a {\it hard-core} in history space 
projecting out certain contributions that would otherwise be allowed by 
the free theory alone. This phenomenon takes the form of a nonstandard, 
nonclassical counterterm in the Hamiltonian that does {\it not} vanish as 
the coupling constant of the quartic interaction vanishes. Specifically, 
for the model in question, the additional counterterm is {\it a counterterm 
for the kinetic energy} and is formally proportional to $\hbar^2/\phi(x)^2$, 
which in form is not unlike the centripetal potential that arises in 
spherical coordinates in three-dimensional quantum mechanics. In summary, 
inclusion of a formal additional interaction proportional to 
$\hbar^2/\phi(x)^2$ in the Hamiltonian density is sufficient to result in 
a well defined and nontrivial (i.e., non-Gaussian) quantum theory for 
interacting ultralocal scalar models. In addition, it may be shown 
\cite{KBOOK} that the classical limit of such quantum theories reproduces 
the classical model with which one started.

The foregoing brief summary holds rigorously for the ultralocal scalar 
fields, and it is conjectured that a suitable counterterm would lead to an 
acceptable intermediate behavior for the relativistic models $\phi^4_n$, 
$n\ge5$. What form should the counterterm take in the case of the 
relativistic $\phi^4_n$ models? We can make a plausible suggestion guided 
by the following general principle that holds in the ultralocal case: The 
counterterm should be an ultralocal (because the kinetic energy is 
ultralocal) potential term arising from the kinetic energy. For the 
relativistic field that argument suggests the counterterm should again 
be proportional to $\hbar^2/\phi(x)^2$. It is also part of this general 
conjecture that the same counterterm is not limited to $\phi^4_n$ models, 
but should be effective for other nonrenormalizable interactions, e.g., 
such as $\phi^6_n$, $n\ge4$, etc. The full argument available at present to 
support this conjecture appears in Chap.~11 of \cite{KBOOK}. (It may even be 
possible to exam this proposal by means of suitable Monte Carlo studies, but 
so far this challenge has not been taken up.)

Note well that the hard-core picture of nonrenormalizable interactions leads 
to such interactions behaving as  {\it discontinuous perturbations}: 
Once turned on, such interactions cannot be completely turned off! Stated 
otherwise, as the nonlinear coupling constant is reduced toward zero, the 
theory passes continuously to a ``pseudofree'' theory---different than the 
``free'' theory---{\it which retains the effects of the hard core}. The 
interacting theory is {\it continuously connected} to the pseudofree theory, 
and may even possess some form of perturbation theory about the pseudofree 
theory. Evidently, the presence of the hard-core interaction makes any 
perturbation theory developed about the original unperturbed theory 
almost totally meaningless.
\subsubsection*{Nonrenormalizable gravity}
Although the differences between gravity and nonrenormalizable scalar 
interaction are significant in their details, there are certain 
similarities we wish to draw on. Most importantly, one can argue 
\cite{KGRAV} that the nonlinear contributions to gravity act as a 
hard-core interaction in a quantization scheme, and thus the general 
picture sketched above for nonrenormalizable scalar fields should apply 
to gravity as well. Assuming that the analogy holds further, there should 
be a nonstandard, nonclassical counterterm that incorporates the dominant, 
irremovable effects of the hard-core interaction. Accepting the principle 
that in such cases perturbation theory offers no clear hint as to what 
counterterms should be chosen, we appeal to the guide used in the scalar 
case. Thus, as our proposed counterterm, we look for an ultralocal 
potential arising from the kinetic energy that appears in the Hamiltonian 
constraint. In fact, the only ultralocal potential that has the right 
transformation properties is proportional to $\hbar^2\s g(x)^{1/2}$. Thus 
we are led to conjecture that the ``nonstandard counterterm'' is none other 
than a term like the familiar cosmological constant contribution! Unlike the 
scalar field which required an unusual term proportional to $1/\phi(x)^2$, 
the gravitational case has resulted in suggesting a term proportional to 
an ``old friend'', namely $g(x)^{1/2}$. At first glance, it seems absurd 
that such a harmless looking term could act to ``save'' the 
nonrenormalizability of gravity. In its favor we simply note that the 
analogy with how other nonrenormalizable theories are ``rescued'' is too 
strong to dismiss the present proposal out of hand---and of course one 
must resist any temptation ``to think perturbatively''. Any attempt to 
consider this possibility must wait until another occasion; we hope to 
return to this subject elsewhere. 

As one small aspect of this problem, let us briefly discuss how the factor 
$\hbar^2$ arises in the gravitational case. Merely from a {\it dimensional} 
point of view, we note that (the first term of) the local kinetic energy 
operator has a formal structure given by
  \bn  -\frac{16\pi G}{c^3}\s\hbar^2 \s\bigg(\s\frac{\delta}{
\delta g_{cb}(x)}\,g_{ac}(x)\,g(x)^{-1/2}\,g_{bd}(x)\,
\frac{\delta}{\delta g_{da}(x)}-\cdots\bigg)\;, \en
where we have restored the factor $16\pi G/c^3$. Thus the anticipated 
counterterm is proportional to $(G\hbar^2/c^3)\s g(x)^{1/2}$. We next 
cast this term into the usual form for a contribution to the potential 
part of the Hamiltonian constraint, namely, in a form proportional to 
$(c^3/G)\s\Lambda\s g(x)^{1/2}$. Hence, to recast our anticipated 
counterterm into this form, we need a factor proportional to
 \bn  \frac{G^2\hbar^2}{c^6}\equiv l_{Planck}^4\approx (10^{-33} cm)^4 \;. \en
In the classical symbol for the Hamiltonian constraint operator, this 
factor is multiplied by an expectation value with dimensions 
${\sf L}^{-6}$ originating from the density nature of the two momentum 
factors and leading to an overall factor with the dimensions 
${\sf L}^{-2}$ that is proportional to $\hbar^2$ as claimed. Let us 
call the resultant counterterm $\Lambda_C\, g(x)^{1/2}$. Since the sign 
of the DeWitt metric that governs the kinetic energy term is indefinite, 
it is not even possible to predict the sign of $\Lambda_C$. However, one 
thing appears certain. While the proposed counterterm 
$\Lambda_C\, g(x)^{1/2}$ is certainly not cosmological in origin, 
its influence may well be!

The foregoing scenario has assumed the hard-core model of nonrenormalizable 
interactions applies to the theory of gravity. However, that may well not 
be the case, and, instead, some other counterterm(s) may be required to cure 
the theory of gravity. Note well that the general structure of our approach 
to quantize gravity is largely insensitive to just what form of 
regularization and renormalization is required. In particular, the 
use of the affine field variables, the application of the projection 
operator method to impose constraints, and the development of the 
nonstandard phase-space functional integral representation for the 
reproducing kernel of the regularized physical Hilbert space all have 
validity quite independently of the form in which the Hamiltonian constraint 
is ultimately turned into a local self-adjoint operator. Although we have 
outlined one particular version in which the Hamiltonian constraint may 
possibly be made into a densely defined local operator, we are happy to 
keep an open mind about the procedure by which this ultimately may take 
place since many different ways in which this process can occur are fully 
compatible with the general principles of our proposed quantization scheme 
for the gravitational field.

\section*{\sc ACKNOWLEDGMENTS}Thanks are expressed to A.~Kempf and G.~Watson 
for comments, and to J.~Govaerts, S.V.~Shabanov, and B.~Whiting for helpful 
suggestions. The work reported in this paper has been partially supported 
by NSF Grant 1614503-12.


\begin{thebibliography}{99}
\bibitem{P-I}J.R.~Klauder, ``Noncanonical Quantization of Gravity. I. 
Foundations of Affine Quantum Gravity'', J. Math. Phys. {\bf 40}, 5860-5882 
(1999), referred to as P-I.

\bibitem{ADM}R.~Arnowitt, S.~Deser, and C.~Misner, in {\it Gravitation: An 
Introduction to Current Research}, Ed. L. Witten, (Wiley \& Sons, New York, 
1962), p. 227.

\bibitem{DIRAC}P.A.M.~Dirac, {\it Lectures on Quantum Mechanics}, (Belfer 
Graduate School of Science, Yeshiva University, New York, 1964).

\bibitem{KORIG}J.R.~Klauder, ``Coherent State Quantization of Constraint 
Systems'', Ann. Phys. {\bf 254}, 419-453 (1997).


\bibitem{UNIV}J.R.~Klauder, ``Universal Procedure for Enforcing Quantum 
Constraints'', Nucl.~Phys.~{\bf B547},  397-412, 1999.

\bibitem{SCHAL}J.R.~Klauder, ``Quantization of Constrained Systems'', 
hep-th/0003297.

\bibitem{SHAB}S.V.~Shabanov, ``Geometry of the Physical 
Phase Space in Quantum Gauge Systems '', Phys. Reports {\bf 326}, 1 (2000).

\bibitem{REL}J.R.~Klauder, ``Ultralocal Fields and their Relevance for
Reparametrization Invariant Quantum Field Theory'', quant-ph/0012076.

\bibitem{HEG}G.C.~Hegerfeldt and J.R.~Klauder, ``Metrics on Test Function 
Space for Canonical Field Operators", Commun. Math. Phys. {\bf 16}, 329-346 
(1970). 

\bibitem{MESH}N.~Aronszajn, Proc. Cambridge Phil. Soc. {\bf 39}, 133 (1943); 
Trans. Am. Math. Soc. {\bf 68}, 337 (1950); H.~Meschkowski, {\it Hilbertsche 
R\"aume mit Kernfunktion}, (Springer Verlag, Berlin, 1962).

\bibitem{12}G.~Watson and J.R.~Klauder, ``Generalized Affine Coherent States: 
A Natural Framework for the Quantization of Metric-like Variables'', J. Math. 
Phys. {\bf 41}, 8072-8082 (2000).

\bibitem{KBOOK}J.R.~Klauder, {\it Beyond Conventional Quantization},
(Cambridge
University Press, Cambridge, 1999).

\bibitem{GUTZ}J.R.~Klauder, ``Coherent State Path Integrals {\it Without} 
Resolutions of Unity'', Found. Phys. (in press), quant-ph/008132.

\bibitem{DAUB}I.~Daubechies and J.R.~Klauder, ``Quantum Mechanical Path 
Integrals with Wiener Measures for All Polynomial Hamiltonians, II", J. Math. 
Phys. {\bf 26}, 2239-2256 (1985); I.~Daubechies, J.R.~Klauder, and T.~Paul, 
``Wiener Measures 
for Path Integrals with Affine Kinematic Variables", {\it J. Math. Phys.} 
{\bf 28}, 85-102 (1987); J.R.~Klauder, ``Quantization {\it Is} Geometry, 
After All", Annals of Physics {\bf 188}, 120-141 (1988). 

\bibitem{CARG}J.R.~Klauder, ``Quantization = Geometry + Probability'', in 
{\it Probabilistic Methods in Quantum Field Theory and Quantum Gravity}, 
Eds. P.H.~Damgaard, H.~H\"uffel, and A.~Rosenblum, (North-Holland, New York, 
1990), p. 73.

\bibitem{MTW}C.~Misner, K.~Thorne, and J.A.~Wheeler, {\it Gravitation}, 
(W.H. Freeman and Co., San Francisco, 1971); R.~Wald, 
{\it General Relativity}, (The University of Chicago Press, Chicago, 1984).

\bibitem{JIMAND}J.~Anderson, in {\it Proceedings of the First Eastern 
Theoretical Physics Conference, University of Virginia, 1962}, Ed. 
M.E.~Rose, (Gordon and Breach, New York, 1963), p.~387.

\bibitem{TEIT}C.~Teitelboim, Phys. Rev. Lett. {\bf 50}, 795 (1983); 
J.J.~Halliwell and J.B.~Hartle, ``Wave Functions Constructed from an 
Invariant Sum-over-histories Satisfy Constraints'', Phys. Rev. D {\bf 43}, 
1170 (1991).

\bibitem{TOBOC}W.~Tobocman, Nuovo cimento {\bf 3}, 12113 (1956).

\bibitem{TMF}S.V.~Shabanov, Theor. Math. Phys. {\bf 78}, 292 (1989); Phys.
Lett. B {\bf 318}, 323 (1993).

\bibitem{SF}F.G.~Scholtz and S.V.~Shabanov, Ann. Phys. (NY) {\bf 263}, 119 
(1998).

\bibitem{ASHT}See, e.g., C. Rovelli, Living Reviews in Relativity, 
http://www.living reviews. org/Articles/Volume1/1998-1rovelli.

\bibitem{PIL}M.~Pilati, Phys. Rev. D {\bf 26}, 2645 (1982); {\bf 28}, 729 
(1983); G.~Francisco and M.~Pilati, {\it ibid.} {\bf 31}, 241 (1985).

\bibitem{SOK}R.~Fern\'andez, J.~Froehlich, and A.~Sokal, {\it Random Walks, 
Critical Phenomena and Triviality in Quantum Field Theory}, (Springer-Verlag, 
New York, 1992).

\bibitem{KGRAV}J.R.~Klauder, ``On the Meaning of a Nonrenormalizable Theory 
of Gravitation", GRG {\bf 6}, 13-19 (1975). 

\end{thebibliography}
\end{document}